# The critical role of the donor polymer in the stability of high-performance non-fullerene acceptor organic solar cells

Yiwen Wang[1], Alberto Privitera[2], Giacomo Londi[3], Alexander J. Sneyd[4], Deping Qian[5], Yoann Olivier[3], Lorenzo Sorace[6], David Beljonne[7], Zhe Li[1*] and Alexander J. Gillett[4*].

[1]School of Engineering and Materials Science, Queen Mary University of London, Mile End Road, London, U.K.

[2]Department of Chemistry, University of Torino, Via Giuria 9, Torino, Italy.

[3]Laboratory for Computational Modeling of Functional Materials, Namur Institute of Structured Matter, Université de Namur, Rue de Bruxelles, 61, 5000 Namur, Belgium.

[4]Cavendish Laboratory, University of Cambridge, JJ Thomson Avenue, Cambridge, U.K.

[5]Department of Chemistry and Centre for Processable Electronics, Imperial College London, 82 Wood Lane, London, UK

[6]Department of Chemistry "Ugo Schiff" and INSTM RU, University of Florence, Via della Lastruccia 3-13, 50019 Sesto Fiorentino, Firenze, Italy.

[7]Laboratory for Chemistry of Novel Materials, Université de Mons, Place du Parc 20, Mons, Belgium.

*Corresponding authors: Alexander J. Gillett: E-mail: ajg216@cam.ac.uk; Zhe Li: E-mail: zhe.li@qmul.ac.uk.

**Driven by the rapid development of non-fullerene electron acceptors (NFAs), the power conversion efficiencies of organic solar cells (OSCs) have reached levels suitable for commercial applications. However, the poor operational stability of high-performance NFA OSCs is a remaining fundamental challenge that must be addressed. Whilst previous studies have primarily focused on the NFA component, we consider here the degradation pathways of both the donor and acceptor materials in the benchmark PM6:Y6 blend. Here, we show that light soaking greatly increases the energetic disorder and trap state density in PM6, with little effect on Y6. This is corroborated by electron paramagnetic resonance spectroscopy, which reveals increased recombination *via* trapped polarons on PM6 after light soaking. In addition, ultrafast optical spectroscopy studies on light-soaked samples show that PM6 singlet excitons are rapidly converted into interchain polaron pairs on sub-100 fs timescales; this process outcompetes electron transfer to Y6, significantly reducing the charge generation yield of the blend. We make similar observations in the parent polymer, PBDB-T, indicating that this class of donor materials, used in most high-performance OSCs to date, are intrinsically unstable to light soaking. Thus, we reveal that the donor polymer can be a further critical weak link in efficient OSC systems, whose degradation mechanism needs to be addressed collectively with NFAs.**

**Introduction**

Through the development of non-fullerene electron acceptors (NFAs), the power conversion efficiencies (PCEs) of organic solar cells (OSCs) have rapidly increased, now exceeding 18%[1–3]. However, whilst the performance of OSC devices is approaching the levels required for their mainstream deployment, such applications also require long-term stability from the photovoltaic modules. Some NFA OSC blends have demonstrated excellent stability, with extrapolated $T_{80}$ lifetimes (time taken for the solar cell PCE to drop to 80% of its initial value) of over 20 years[4,5], but this has not yet been simultaneously achieved in systems which currently give the best PCEs in NFA OSCs to date[6]. Therefore, successful commercialisation necessitates that substantial effort be dedicated to understanding and engineering out the causes of instability in high-performance NFA OSC systems, such as those based on PM6:Y6 and their derivatives[6,7]. When considering the degradation mechanisms of an OSC device, it is important to note that performance losses may result from a complex combination of factors[8–10]. This can include not only an innate instability in the materials or morphology of the polymer:NFA blend[11–13], but also chemical degradation of the blend induced by interlayers in the device stack[14–16], or degradation of the device stack itself[17]. Whilst the latter two issues can be remedied through careful device engineering, including appropriate selection of the interlayers and encapsulation materials[4,5,18–20], any intrinsic instability in the blend components presents a more fundamental challenge that must be addressed separately.

We report here on the intrinsic degradation mechanisms in the light absorbing layers of efficient NFA OSCs. We have chosen the benchmark PM6:Y6 system to act as a representative example (chemical structures in Figure 1a)[7], as many current high-performance OSCs comprise of a 'Y-series' NFA paired with the PM6 donor polymer[2,6,21,22]. In our model study, we selectively age the PM6:Y6 active layer by light soaking, resulting in a significant decrease in the PCE of the PM6:Y6 device. By combining optical and magnetic resonance spectroscopies with device stability studies, we primarily attribute this loss in performance to degradation of the donor polymer PM6, not the NFA Y6. We observe similar behaviour upon light soaking the 'parent' material, PBDB-T, suggesting that these loss mechanisms are intrinsic to this polymer class. Thus, whilst the field is heavily focussed on advancing the design of NFAs, we propose that the donor polymer design is equally important and should not be overlooked when considering strategies to improve the stability of high-performance NFA OSCs.

**Fresh and light-soaked solar cell characteristics**

To investigate the degradation mechanisms intrinsic to the PM6:Y6 blend, we have fabricated inverted devices with the structure of ITO/ZnO/PM6:Y6/MoO$_3$/Ag. To ensure degradation is limited to, as far as possible, the PM6:Y6 active layer, we have deposited the PM6:Y6 film on top of the ITO/ZnO layers, before light soaking the samples under 1 Sun equivalent white LED illumination. Following light soaking, fresh layers of MoO$_3$ and Ag are evaporated on top of the bulk heterojunction (BHJ) layer to complete the device. For brevity, we focus on films light soaked in ambient air but note that similar behaviour is observed for light soaking in a nitrogen environment (*vide infra*). Thus, whilst oxygen and water are likely to play an additional role in accelerating device degradation, here, we primarily explore the effect of light soaking.

In Figure 1b, we present the current density-voltage characteristics of the fresh PM6:Y6 devices and devices where the BHJ was selectively light soaked for 0.5, 3, 6, and 12 hours (photovoltaic parameters are summarized in Table 1). Fresh devices show a PCE of 14.9%, with a short circuit current density ($J_{SC}$) of 26.99 mA/cm$^2$, an open-circuit voltage ($V_{OC}$) of 0.80 V and a fill-factor (FF) of 68.8%. Upon light soaking, the device performance decreases severely, indicating significant degradation of the PM6:Y6 BHJ. The external quantum efficiency (EQE) spectrum of the fresh and aged blends is substantially reduced across the whole spectrum after light soaking (Figure 1c). However, the region between 500-700 nm that corresponds to the PM6 absorption shows a stronger decrease, suggesting that charge generation pathways from excitons created on PM6 are more affected than those from Y6.

To probe the charge carrier recombination mechanisms, we have investigated the impact of light intensity on the device parameters (results are summarised in Table S1). In Figure S1a, we plot $V_{OC}$ vs the natural logarithm of the light intensity (ln $I$). Fresh devices have a gradient of $1.00 \cdot \frac{kT}{q}$, confirming that non-geminate recombination is the primary recombination process[23]. In contrast, a gradient of $1.37 \cdot \frac{kT}{q}$ is obtained for the 3-hour light soaked PM6:Y6 OSCs, closer to $2 \cdot \frac{kT}{q}$. Here, the steeper gradient of the 3-hour light-soaked device illustrates that trap-mediated recombination is increased upon aging; for longer light soaking timescales, the diode behaviour of the device begins to break down and anomalous gradients are obtained. The $J_{SC}$ vs (ln $I$) characteristics, measured for the fresh and aged devices, are plotted in Figure S1b. The $J_{SC}$ is dependent on the intensity of the incident light ($J_{SC} \propto I^{\beta}$), where the exponent $\beta$ is typically between 0.75 and 1; a deviation of $\beta$ below unity is closely associated with non-geminate recombination in the device[24,25]. We obtain $\beta \sim 0.92$ for the fresh, 0.5-, and 3-hour light-soaked devices, indicating that there is little change in the non-geminate recombination in the PM6:Y6 blend under these aging timescales. Thus, the changes in the device parameters after light soaking suggest that the PM6:Y6 blends suffers from more trap-mediated recombination but little change in the bimolecular recombination.

**Steady-state optical characterisation**

To understand the degradation mechanisms taking place in the PM6:Y6 active layer, we begin by characterising the films using steady-state techniques. In Figure 2a, we show the UV-Vis absorption spectrum of fresh and aged neat PM6 and Y6 films, with the corresponding neat and blend films aged in air and nitrogen displayed in the SI (Figures S2-S3). To demonstrate most clearly the impact of light-soaking, we focus herein on films light-soaked for 12 hours where the greatest degree of degradation is expected to occur. We observe no significant change in the absorption spectrum of the neat Y6 film (Figure 2a) after light-soaking, and the photoluminescence quantum efficiency (PLQE) effectively remains constant within the experimental error of ±5%: 2.0% in the fresh and 2.1% in the aged film (emission spectra shown in Figure S4). However, the absorption spectrum of PM6 clearly changes after light-soaking for 12 hours. Whilst there is only a small decrease in the absorption magnitude of the PM6 band centred at 600 nm upon aging, the vibronic structure of this feature is lost in the neat film (Figure 2a) and PM6:Y6 blend (Figure S2a). We also note an increased intensity in the PM6 absorption tail, potentially indicating a larger number of sub band gap states. In addition to the absorption changes, we find that the PLQE drops from 0.3% in the fresh PM6 film to below the detection limit of our experimental setup (<0.1%) in the aged film (emission spectra shown in Figure S4), signifying an enhancement of the non-radiative processes quenching bright singlet excitons ($S_1$) in PM6.

To probe the PM6 absorption changes further, we have performed highly sensitive photothermal deflection spectroscopy (PDS) absorption measurements on fresh and aged neat PM6 films, as well as fresh and aged PM6:Y6 blend films (Figure 2b). PDS reveals a significant increase in the Urbach energy of neat PM6 after light-soaking to 84.8 meV, compared to 49.3 meV in the fresh film. This observation, along with the loss of the vibronic peak progression in the PM6 absorption spectrum (clear vibronic features are associated with a higher degree of ordering between the polymer chains[26]), confirm that light-soaking increases the energetic disorder in the PM6 domains[27,28]. In contrast, the PDS absorption of the fresh and aged PM6:Y6 blends have an identical Urbach energy for the Y6 absorption tail of 29.5 meV. We therefore conclude that the primary changes in the blend upon light-soaking occur in the PM6 domains. In addition, we also observe an increase in the intensity of the sub band gap absorption feature around 1 eV in both the light soaked neat PM6 and PM6:Y6 blend films; as this band is present in the neat PM6 film, we determine that it represents an increased density of PM6 trap states[27–29]. We note that similar low energy absorption bands in neat polymer films have been attributed to direct optical transitions associated with interchain polaron pair (PP) states[30], potentially suggesting the nature of these sub gap trap species (*vide infra*).

**Transient absorption studies**

To better understand how the sample degradation is linked to the loss in photovoltaic performance, we have studied the fresh and aged PM6:Y6 films using transient absorption (TA) spectroscopy. This technique can provide an unparalleled insight into both optically bright and dark states over a range of timescales spanning from femtoseconds to milliseconds, allowing for the direct observation of the charge transfer processes between the donor polymer and the NFA[31–35], separation of the intermediate charge transfer state into free charges[36–38], and the charge carrier recombination mechanisms[39–41]. Thus, TA is ideally suited to provide information on the impact of blend degradation on these critical photophysical processes that underpin photovoltaic operation. However, despite the powerful insights that can be gained, TA is not frequently utilised in conjunction with OSC degradation studies to identify loss mechanisms in the blend itself[29,42,43].

Consistent with the steady-state measurements, we find no significant differences between the photophysics of fresh and aged neat Y6 films (Figure S5). In the corresponding

PM6:Y6 blends, with selective excitation of Y6 at 800 nm (Figure S6), we observe no change in the intrinsic hole transfer rate, but a slightly slower quenching of excitons generated on Y6 in the aged blend. As a result, there is a small reduction in the charge yield of the aged PM6:Y6 blend following excitation of Y6. These observations are consistent with the donor:acceptor (D:A) de-mixing previously observed in PM6:'Y-series' NFA blends[12], which increases the average distance that excited states must diffuse to the D:A interface and will lead to a larger fraction of excitations on Y6 recombining prior to hole transfer.

As our previous results suggest that PM6 is more affected by light soaking than Y6, we primarily focus on the photophysics of the donor polymer. We begin with the TA of the fresh PM6 film (Figure 3a). After excitation at 580 nm, we observe the PM6 ground state bleach (GSB) between 550-650 nm, a weaker PM6 stimulated emission (SE) band at 720 nm, and a broad photo-induced absorption (PIA) extending into the near infrared, with a peak at 1150 nm. The SE band and PIA at 1150 nm provide clear fingerprints for the presence of PM6 $S_1$ states. Interestingly, we observe the decay of the PM6 SE and $S_1$ PIA on picosecond timescales, accompanied by the formation of a new band at 920 nm. Through comparison to the TA of a reference PM6:PC$_{60}$BM blend (Figure S7), we assign this feature to the PM6 hole polaron. We therefore conclude that in the fresh PM6 film, $S_1$ states convert to interchain PPs[44]; this has been observed in polymer films where close contact between the individual chains allows for interchain charge transfer to take place[30,45–47]. However, given the short PM6 GSB lifetime (Figure 3c) and low PLQE of the fresh PM6 film (0.3%), we conclude that the PP states formed are tightly bound and do not achieve long range charge separation, resulting in rapid non-radiative recombination.

We next turn to the TA of the aged PM6 film (Figure 3b), also excited at 580 nm. Despite the similar absorbance of the fresh and aged PM6 films at 580 nm (Figure 2a), the intensity of the PM6 $S_1$ PIA at 1150 nm is a factor of two lower in the aged film at 0.1-0.15 ps after excitation with the same fluence (Figure 3d). The resolvable decay kinetic of the $S_1$ PIA after 0.1 ps also shows a more rapid decrease in the aged film. In addition, the PM6 SE band is almost completely quenched by 0.1-0.15 ps, whilst the hole polaron PIA at 920 nm is nearly fully formed at this time. The observations we make here for the neat PM6 film light soaked in air are very similar to the behaviour in the neat PM6 film light soaked in a nitrogen environment (Figure S8), demonstrating that light soaking is the primary factor driving these changes. Thus, we conclude that the rate of interchain PP formation in PM6 is significantly increased upon light soaking, with a substantial fraction of the PPs formed on sub-100 fs timescales; this is consistent with the unmeasurable PLQE (<0.1%) of the aged PM6 film, as nearly all the PM6 $S_1$ states are quenched before they can emit. To explain faster rate of PP formation in aged PM6, we note that an increased Urbach energy is related to a larger number of trap sites[27–29]. Therefore, we propose that the greatly increased Urbach energy of the aged PM6 film (Figure 2b) reflects a higher trap density in PM6 domains, meaning that PM6 $S_1$ states have a greater chance of being generated near a trap site favourable for the formation of bound PP states[48].

In the TA of the fresh and aged PM6:Y6 blend films with preferential excitation of PM6 at 580 nm (Figure S9), we find that the interchain PP formation is fast enough to out-compete electron transfer to Y6 (see SI for detailed discussion). Furthermore, as the interchain PPs in PM6 have a much shorter lifetime and slower transport through the PM6 domains than $S_1$ states[44], they are susceptible to recombining before reaching the D:A interface for electron transfer. Indeed, we observe a divergence in the PM6 GSB kinetics of the fresh and aged blend films on the same tens of picoseconds timescales that the interchain PP states recombine over (Figure S9c). Therefore, as the PM6 $S_1$ states are more rapidly funnelled through this channel in the aged samples, this results in increased losses though interchain PP recombination. The impact of the PP loss channel is clearly seen in the nanosecond TA measurements of the fresh and aged blend films, where the intensity of the PM6 hole polaron signal at 1 ns in the aged blend is about half that of the fresh blend (Figure S10). This difference in the hole polaron yield also agrees well with the roughly factor of two lower intensity of the PM6 $S_1$ PIA at 100 fs in the aged PM6 and PM6:Y6 blend film TA (Figures 3d and S9d), indicating that the charge yield following the electron transfer process is severely limited by the PM6 interchain PP loss pathway.

To understand whether the interchain PP loss pathway is intrinsic to donor polymers originating from the PBDB-T class of materials, we have also investigated the parent compound, PBDB-T. The UV-Vis absorption spectra of neat PBDB-T film before and after light soaking for 12 hours in air are plotted in Figure S14. The PBDB-T film absorption exhibits the same degradation changes as PM6, with a clear loss of the vibronic structure in the absorption band centred at 600 nm and an increase in the intensity of the absorption tail. In the TA of fresh and aged PBDB-T films (Figure S15), we again observe similar behaviour with an enhanced formation rate of the PBDB-T hole polaron PIA in the aged sample. Thus, we confirm that the increased disorder and enhanced PP formation upon light soaking appears to be intrinsic to this family of polymers.

**Light-induced electron paramagnetic resonance studies**

To investigate the degradation mechanisms over a longer photophysical timescales (seconds) and to clarify the nature of the long-lived species generated by light, such as trapped polarons, we turn to light-induced electron paramagnetic resonance (L-EPR) spectroscopy. As a steady-state technique, the L-EPR signal intensity is determined by the balance between the photogeneration efficiency and the polaron recombination timescales[49–51]. Thus, organic thin films with efficient charge generation and slow recombination (e.g. *via* trap states) are expected to display strong signals. In Figures 4a and 4b, we show the L-EPR spectra of the fresh and aged neat PM6 and Y6 films, and the PM6:Y6 blend (T = 40 K). Consistent with the TA studies, we have excited the films at 534 and 785 nm for preferential excitation of PM6 and Y6, respectively. The fresh and aged neat Y6 films do not show any appreciable signal which implies that the photogeneration efficiency is low. However, whilst the fresh neat PM6 film also does not show any signal, a weak L-EPR response is observed in the aged PM6 film. The presence of a L-EPR signal in the aged PM6 film cannot be attributed to differences in the polaron generation efficiencies between the fresh and aged PM6 films, as they behave similarly in the TA beyond the initial PP generation dynamics in the first few picoseconds (Figure 3). Thus, we conclude that this signal is due to the presence of longer-lived species, which we attribute to more deeply trapped polarons with slower recombination in the aged PM6 film. In the PM6:Y6 blends, a narrow and almost isotropic L-EPR signal with g = 2.0028 ±0.0005 is observed, characteristic of long-lived photogenerated polarons in organic semiconductors[50]. The g-value and line shape of this feature are very similar to those of the signal observed in the aged PM6 film, suggesting that this species is localised on PM6. Indeed, quantum-chemical calculations provide isotropic g-factors of 2.0029 and 2.0032 for the polarons of PM6 and Y6 respectively (see Methods for details), confirming the attribution of the observed signal to the PM6 polaron. The best-fit simulations of this feature reveals that the PM6 polaron linewidth is broader in the aged blend (Figure S16, Table S2). This result confirms a wider range of polaron environments in the aged blend compared to the fresh blend and is indicative of a wider trap energy distribution in the aged PM6:Y6 sample[52]. The double integration of the L-EPR signals after 534 nm excitation further reveals that the signal intensity of the fresh blend is roughly a factor of two more intense (Figure S17), consistent with observations in the TA (Figure S11). This indicates that the significantly higher photogeneration efficiency of the fresh PM6:Y6 blend overshadows the deeper traps in the aged blend, ultimately resulting in a stronger L-EPR signal.

To obtain further insights into the trap distribution and recombination processes, we record the decay of the PM6 polaron EPR signal in the fresh and aged PM6:Y6 blends at 10 K after the light source is turned off (Figures 4c and 4d). Whilst the rapid growth of the L-EPR signal after 534 nm illumination at -30 s indicates the expected rapid polaron photogeneration, the recombination kinetics of both samples reveals a signal that decays over longer timescales, indicating the presence of trapped polarons[53–55]. The model typically employed to explain this slow polaron recombination is one of 'trap-limited bimolecular recombination', where trapped polarons can only recombine after thermally activated detrapping[56]. The bi-exponential fit of the signal decay indicates longer recombination timescales in the aged PM6:Y6 blend than in the fresh blend. This result corroborates the presence of a more defective environment with a higher density of deep PM6 traps in the aged blend, in line with the increased disorder in PM6 domains upon aging (Figure 2). We also note a clear change in the recombination mechanism in the nanosecond TA measurements of PM6:Y6 (Figure S11), with the PM6 polaron recombination exhibiting a much weaker fluence dependence in the light-soaked blend. This observation further supports a shift from bimolecular recombination in the fresh blend to more dominant trap mediated recombination in the aged blend.

**Conclusions**

In summary, we have investigated the degradation pathways intrinsic to the benchmark PM6:Y6 system through model stability studies and subsequent spectroscopic analysis. Whilst Y6 appears to be robust under light soaking, PM6 shows significantly increased energetic disorder upon aging, as evidenced through steady-state absorption and PDS spectroscopies. The increased disorder in PM6 domains induced by light soaking leads to the rapid formation of trapped PP states on sub-100 fs timescales, severely limiting the charge generation in the device. Furthermore, the increased disorder in PM6 results in a more defective environment and increased recombination *via* trapped polarons on PM6. Thus, we conclude that a key factor limiting the stability of high performance OSCs comprised of PM6 blended with a 'Y-series' NFA is the donor polymer, PM6. We make similar observations in the 'parent' polymer of PM6, PBDB-T, indicating that this degradation pathway is more broadly present in this class of donor materials. We note that many previous studies on the stability of NFA OSC blends have focussed on the NFA component[10,57], with less attention given to the donor polymer. Our study therefore shows the critical and often overlooked importance of the polymer to the degradation of OSCs, which needs to be understood and addressed collectively with that of the NFA component.

Furthermore, as the majority of high-performance (PCE >15%) OSC systems reported to date are based on PM6 blended with a 'Y-series' NFA[2,3,6,7], our results have important consequences for the future development of OSCs. Namely, if high performance and excellent stability are to be simultaneously achieved, the field must move beyond using PM6 and other polymers from this material family as electron donors. To this end, it is important to conduct stability studies on other polymers that have demonstrated good performance with 'Y-series' NFAs, such as D18 and PTQ10[1,58], in order to identify donor materials that could enable improved device stability. However, the promising photostability of Y6 indicates that if more stable donor polymers can be identified, NFA OSCs have the potential to simultaneously demonstrate the high performance and long operational lifetimes necessary for widespread commercial utilisation.

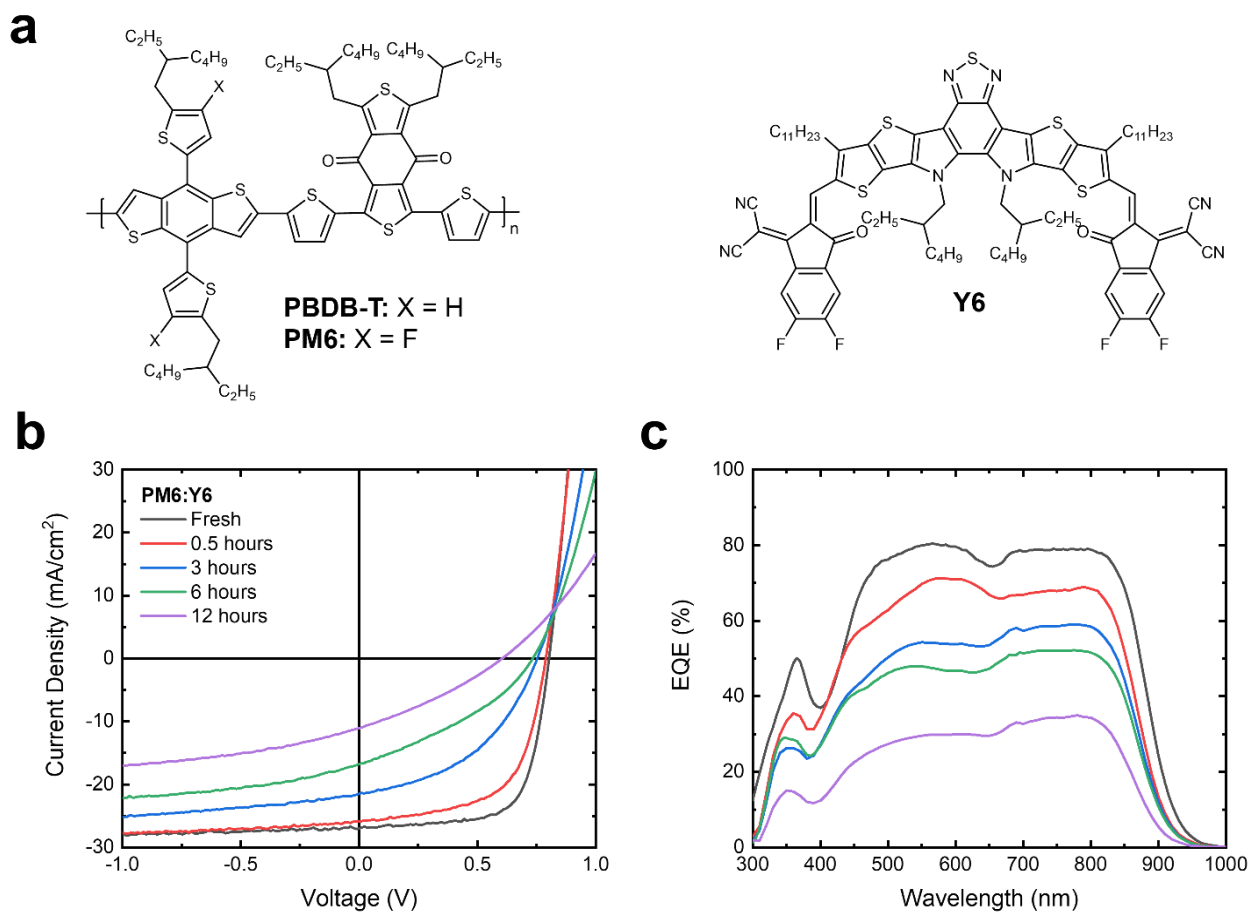

**Figure 1: The chemical structures and solar cell performance of the materials investigated in this study** (**a**) The chemical structures of the donor polymers, PBDB-T and PM6, and the acceptor material Y6, explored in this study. (**b**) The current density-voltage curves of the PM6:Y6 solar cells where the BHJ layer was light-soaked for varying timescales under 1 Sun intensity white LED illumination in ambient air. Device measurements were performed under 100 mW/cm$^2$ AM1.5G solar illumination. (**c**) The corresponding external quantum efficiency (EQE) curves of the aged PM6:Y6 solar cells presented in Figure 1b.

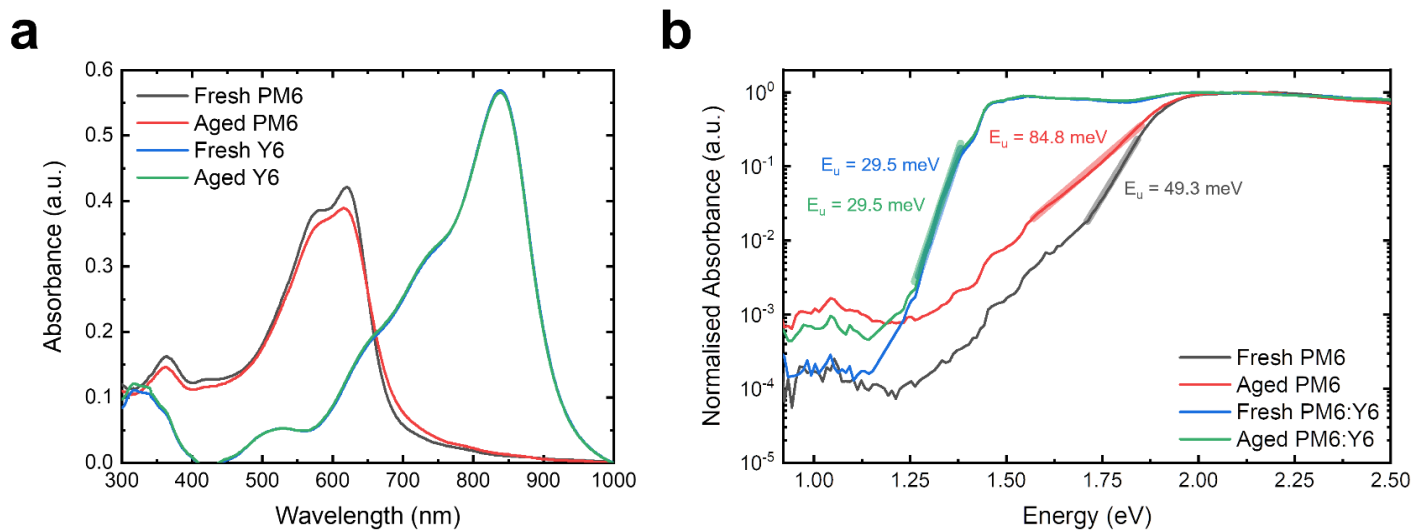

**Figure 2: The steady-state absorption spectra of fresh and aged PM6 and Y6 films (a)** The UV-Vis spectra of neat PM6 and Y6. Absorption spectra for both fresh films and samples light-soaked for 12 hours in ambient air are shown. **(b)** The absorption spectra of PM6 and the PM6:Y6 blend measured with photo-thermal deflection spectroscopy. Spectra for both fresh films and samples light-soaked for 12 hours in ambient air are shown. The Urbach energy, obtained from fitting the absorption tail of the samples (fit indicated by the thick semitransparent line), is also included on the plot.

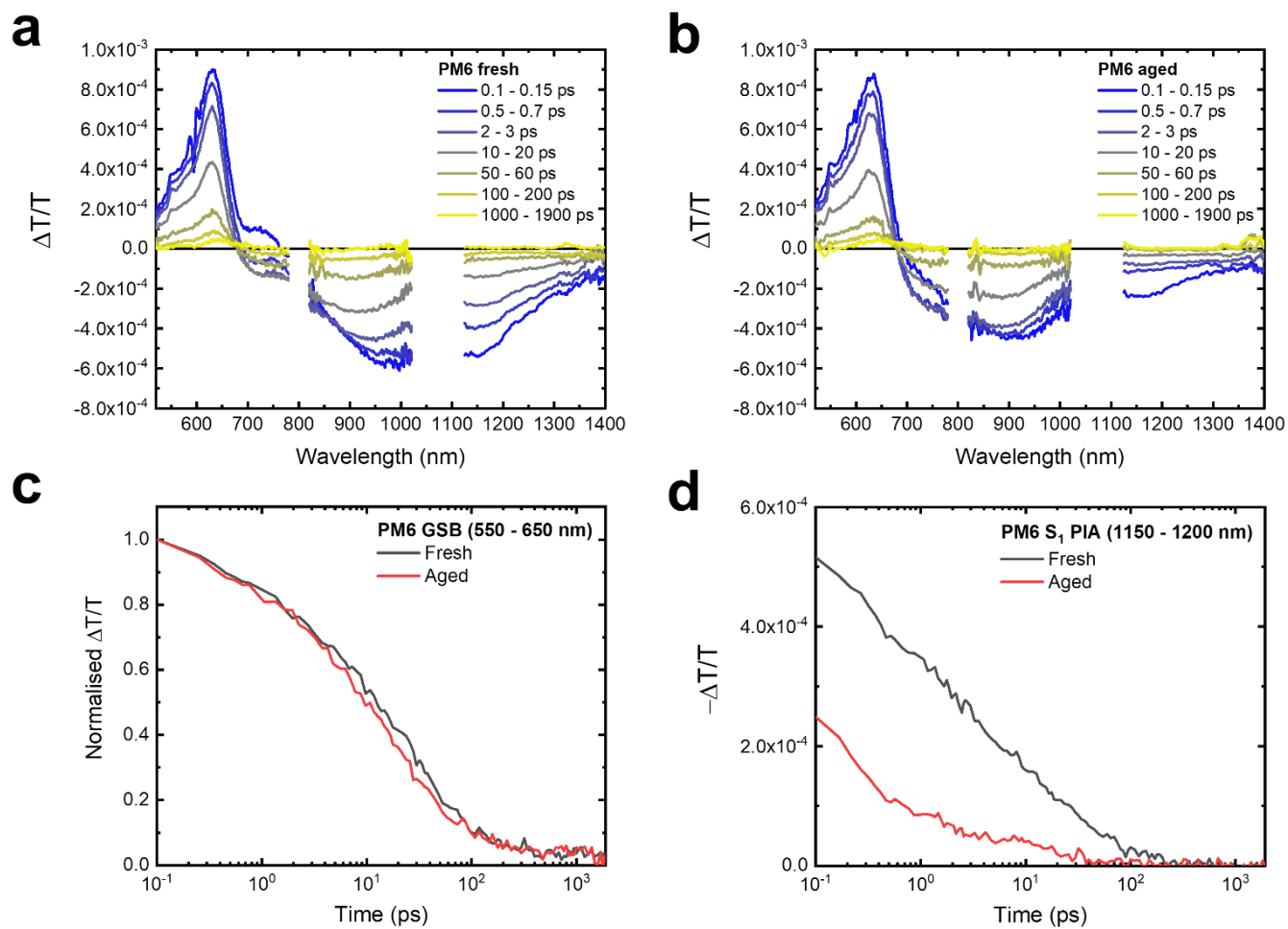

**Figure 3: The TA spectra and kinetics of fresh and aged PM6 films.** (**a**) The TA spectra of the fresh neat PM6 film, excited at 580 nm with a fluence of 0.64 μJ/cm$^2$. (**b**) The TA spectra of the neat PM6 film, light soaked for 12 hours in ambient air, excited at 580 nm with a fluence of 0.64 μJ/cm$^2$. (**c**) The normalized kinetics of the PM6 GSB region (550-650 nm), taken from the fresh and aged neat PM6 films. (**d**) The raw kinetics of the PM6 S$_1$ PIA (1150-1200 nm), taken from the fresh and aged PM6 films.

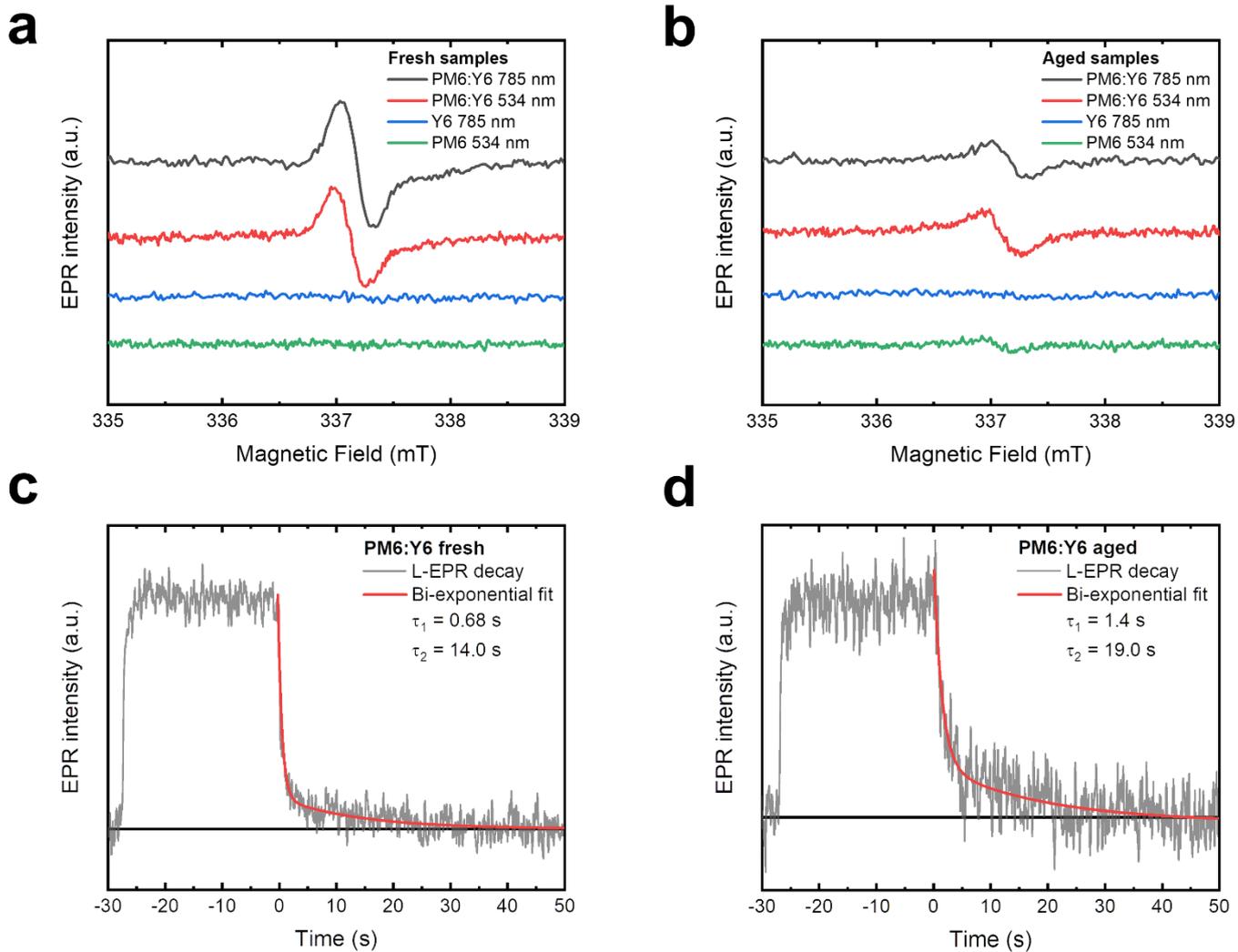

**Figure 4: The light-induced EPR of fresh and aged samples.** (a) The L-EPR spectra (T = 40 K) of fresh PM6 and Y6 neat films, as well as a fresh PM6:Y6 blend film excited at 534 and 785 nm. (b) The L-EPR spectra (T = 40 K) of aged PM6 and Y6 neat films, as well as an aged PM6:Y6 blend film excited at 534 and 785 nm. (c) The L-EPR transient (T = 10 K) of the fresh PM6:Y6 film with 534 nm excitation, taken at a field position of 337 mT, corresponding to the maximum of the EPR line. Illumination was turned off at t = 0 s. (d) The L-EPR transient (T = 10 K) of the aged PM6:Y6 film with 534 nm excitation, taken at a field position of 337 mT. Illumination was turned off at t = 0 s.

| Light soaking time | PCE (%) | $V_{OC}$ (V) | $J_{SC}$ (mA/cm$^2$) | FF (%) |
|---|---|---|---|---|
| Fresh | 14.87 | 0.80 | 26.99 | 68.77 |
| 0.5 hours | 12.38 | 0.79 | 25.81 | 60.89 |
| 3 hours | 7.30 | 0.75 | 21.50 | 45.04 |
| 6 hours | 4.27 | 0.73 | 16.75 | 34.73 |
| 12 hours | 2.05 | 0.60 | 11.08 | 30.65 |

**Table 1: The device performance parameters of the fresh and aged PM6:Y6 solar cell devices.**

## Methods

### OSC device fabrication

PM6 and Y6 were purchased from Solarmer. Neat PM6 and Y6 solution was prepared in chloroform at 8 mg mL$^{-1}$ and 10 mg mL$^{-1}$, respectively. PM6:Y6 (1:1.2) blends were dissolved in chloroform at 18 mg mL$^{-1}$ with 0.5% chloronaphthalene (CN). All organic solutions were stirred at 40 °C for at least 4 hours before use. Devices were fabricated with the structure of ITO/ZnO/PM6:Y6/MoO$_3$/Ag. The substrates were cleaned by ultrasonication sequentially with acetone, dilute detergent solution, deionized water, acetone, and isopropanol for 20 min each before being dried under a stream on N$_2$ and treated with UV–ozone immediately prior to use. ZnO sol-gel was then spin-coated on the ITO substrates and annealed at 150 °C for 10 min in air. After transferring samples to the glovebox with O$_2$ and H$_2$O levels <0.1 ppm, a layer of PM6:Y6 (80-100 nm) was spin-coated on the ZnO/ITO surface. The samples were then light soaked for varying timescales (0.5h, 3h, 6h and 12 h) under 1 Sun intensity white LED illumination in ambient air or N$_2$. Then, the samples were transferred to the adjacent vacuum chamber with a base pressure of <2.0×10$^{-6}$ Pa for deposition of a 10 nm thick MoO$_3$ interlayer and a 100 nm thick Ag top electrode. The OSCs have an identical active area of 0.045 cm$^2$, defined by the overlap area of the anode and the cathode. The active area of the device was determined using an optical microscope.

### OSC device testing

J-V characteristics of fresh and aged OSC devices were measured under AM1.5G illumination of 100 mw/cm$^2$ using a xenon lamp solar simulator in an inert environment. Eight individual solar cell devices were tested for each aging timescale reported. A Keithley 2635A source measurement unit was used to scan the voltage applied to the solar cell between -2 to 1 V at a speed of 0.43 V/s with a dwell time of 46 ms. Light intensity of the solar simulator was calibrated using a single crystalline silicon detector (with a KG-5 filter) to minimize the spectral mismatch. A Bentham PVE300 photovoltaic QE system was used to obtain the EQE spectrum over the wavelength range 300–1000 nm.

**TA spectroscopy**

TA was performed on a setup powered using a commercially available Ti:sapphire amplifier (Spectra Physics Solstice Ace). The amplifier operates at 1 kHz and generates 100 fs pulses centred at 800 nm with an output of 7 W. A TOPAS optical parametric amplifier (OPA) was used to provide the tuneable ~100 fs pump pulses for the picosecond TA measurements. For the nanosecond TA measurements, the second harmonic (532 nm) of an electronically triggered, Q-switched Nd:YVO4 laser (Innolas Picolo 25) provided the ~1 ns pump pulse. The probe pulses were provided by broadband non-collinear OPAs (NOPAs) operating in the visible and near infrared spectral regions. The probe pulses are collected with an InGaAs dual-line array detector (Hamamatsu G11608-512DA), driven and read out by a custom-built board from Stresing Entwicklungsbüro. The probe beam was split into two identical beams by a 50/50 beamsplitter. This allowed for the use of a second reference beam which also passes through the sample but does not interact with the pump. The role of the reference was to correct for any shot-to-shot fluctuations in the probe that would otherwise greatly increase the structured noise in our experiments. Through this arrangement, very small signals with a $\frac{\Delta T}{T} = 1 \times 10^{-5}$ could be measured.

**Photoluminescence quantum efficiency measurements**

The PLQE was determined using method previously described by De Mello *et al.*[59]. Samples were placed in an integrating sphere and photoexcited using a 658 nm continuous-wave laser.

The laser and emission signals were measured and quantified using calibrated Andor iDus DU420A BVF Si and Andor CCD-1430 InGaAs detectors.

**L-EPR spectroscopy**

Light-induced EPR measurements were performed at X-band ($\nu \sim 9.40$ GHz) on a Bruker Elexsys E500 spectrometer, equipped with a $^4$He continuous flow cryostat (ESR 900, Oxford Instruments) for measurements at variable temperatures (10 and 40 K). All the spectra were acquired by using the same modulation frequency (100 kHz), modulation amplitude (1 G), microwave power (0.02 mW, 40 dB) and receiver gain (60 dB). The reported spectra were averaged 49 times. The magnetic field was calibrated with a crystal of DPPH. For sample illumination, Thorlabs laser diodes ($\lambda \sim 534$ and 785 nm) were focused in front of the optical access of the EPR spectrometer cavity (Bruker ST102). The system delivered about 35 and 50 mW of light power to the sample at 534 and 785 nm, respectively. The power was measured at the output of the laser diode, before the EPR cavity grid. EPR spectral simulations were carried out using the core functions *pepper* and *esfit* of the open-source MATLAB toolbox EasySpin[60].

EPR samples were fabricated by spin-coating solutions under identical conditions to the optimised devices onto Mylar substrates, which were subsequently cut into strips with a width of 3 mm. To ensure the flexible Mylar substrates did not bend during the spin coating process, they were mounted onto rigid glass substrates using adhesive tape. Ten strips were placed in quartz EPR tubes; the same amount of material in each sample ensures that EPR intensities can be compared for the equivalent fresh and aged neat or blend film samples excited with the same wavelength. The EPR tubes were sealed in a nitrogen glovebox with a mono-component resin (SIGILL), ensuring that all EPR measurements were performed without oxygen exposure.

**Photothermal deflection spectroscopy**

PDS sensitively measures absorption directly by probing the heating effect in samples upon absorption of light. Films were coated on a Spectrosil fused silica substrate and were immersed in an inert liquid FC-72 Fluorinert. They were then excited with a modulated monochromated light beam perpendicular to the plane of the sample. A combination of a Light Support MKII 100 W Xenon arc source and a CVI DK240 monochromator was used to produce a modulated monochromated light beam. The PDS measurements were acquired by monitoring the deflection of a fixed wavelength (670 nm) diode laser probe beam following absorption of each monochromatic pump wavelength.

**Quantum-chemical calculations**

The ground-state geometry optimizations of an isolated PM6 trimer and the Y6 NFA were performed at the DFT level, using the ωB97X-D functional and the 6-31G(d,p) basis set. All the alkyl chains were replaced by methyl groups to speed up the DFT optimizations, carried out with the Gaussian16 software[61]. Then, magnetic properties of the charged species (*i.e.*, the PM6 cation and the Y6 anion) were computed at the B3LYP/Def2-TZVP level of theory. Namely, g-tensor calculations were performed by using gauge including atomic orbitals (GIAOs[62]) and the RIJCOSX approximation. These calculations were done using the ORCA 4.2 code[63].

# Data availability

The data that support the plots within this paper is available at the University of Cambridge Repository: [to be completed in proofs].

# Acknowledgements

AJG acknowledges support from the EPSRC (EP/M01083X/1 and EP/M005143/1), and the Strategic University Network to Revolutionise Indian Solar Energy (SUNRISE), EPSRC grant ref. EP/P032591/1. This project has received funding from the ERC under the European Union's Horizon 2020 research and innovation programme (grant agreement no. 670405). YW and ZL acknowledge the funding of UK Engineering and Physical Sciences Research Council (EP/S020748/1 and EP/S020748/2). AP and LS acknowledge support from the Italian Ministry of Education and Research (MIUR) through PRIN project 2017 "Quantum detection of chiral-induced spin selectivity at the molecular level" (2017Z55KCW). Computational resources were provided by the Consortium des Équipements de Calcul Intensif (CÉCI), funded by the Fonds de la Recherche Scientifiques de Belgique (F.R.S.-FNRS) under Grant No. 2.5020.11, as well as the Tier-1 supercomputer of the Fedération Wallonie-Bruxelles, infrastructure funded by the Walloon Region under Grant Agreement No. 1117545. GL and YO acknowledge funding by the Fonds de la Recherche Scientifique-FNRS under Grant n° F.4534.21 (MIS-IMAGINE). DB is a FNRS Research Director. We thank Richard Friend for helpful discussions.

## Author contributions

AJG, YW, and ZL conceived the work. AJG performed the TA and PLQE measurements. YW characterised the OSC devices, prepared the samples for the spectroscopy studies, and performed the steady-state absorption measurements. AP conducted the L-EPR studies. GL, YO, and DB carried out the quantum-chemical calculations. AJS measured the PDS. DQ performed the photoluminescence measurements. YO, LS, DB and ZL supervised their group members involved in the project. AJG, YW, and ZL wrote the manuscript with input from all authors.

## Competing interests

The authors declare no competing interests.

## Additional information

Supplementary information accompanies this paper at [to be completed in proofs].

Correspondence and requests for materials should be addressed to AJG (ajg216@cam.ac.uk) and ZL (zhe.li@qmul.ac.uk).

# References

1   Q. Liu, Y. Jiang, K. Jin, J. Qin, J. Xu, W. Li, J. Xiong, J. Liu, Z. Xiao, K. Sun, S. Yang, X. Zhang and L. Ding, *Sci. Bull.*, 2020, **65**, 272–275.

2   Y. Cai, Y. Li, R. Wang, H. Wu, Z. Chen, J. Zhang, Z. Ma, X. Hao, Y. Zhao, C. Zhang, F. Huang and Y. Sun, *Adv. Mater.*, 2021, **33**, 2101733.

3   P. Bi, S. Zhang, Z. Chen, Y. Xu, Y. Cui, T. Zhang, J. Ren, J. Qin, L. Hong, X. Hao and J. Hou, *Joule*, 2021, **5**, 2408–2419.

4   X. Xu, J. Xiao, G. Zhang, L. Wei, X. Jiao, H.-L. Yip and Y. Cao, *Sci. Bull.*, 2020, **65**, 208–216.

5   Y. Li, X. Huang, K. Ding, H. K. M. Sheriff, L. Ye, H. Liu, C.-Z. Li, H. Ade and S. R. Forrest, *Nat. Commun.*, 2021, **12**, 5419.

6   S. Li, C. Li, M. Shi and H. Chen, *ACS Energy Lett.*, 2020, **5**, 1554–1567.

7   J. Yuan, Y. Zhang, L. Zhou, G. Zhang, H.-L. Yip, T.-K. Lau, X. Lu, C. Zhu, H. Peng, P. A. Johnson, M. Leclerc, Y. Cao, J. Ulanski, Y. Li and Y. Zou, *Joule*, 2019, **3**, 1140–1151.

8   W. Li, D. Liu and T. Wang, *Adv. Funct. Mater.*, 2021, **2104552**, 2104552.

9   Q. Burlingame, M. Ball and Y.-L. Loo, *Nat. Energy*, 2020, **5**, 947–949.

10  Y. Wang, J. Lee, X. Hou, C. Labanti, J. Yan, E. Mazzolini, A. Parhar, J. Nelson, J. Kim and Z. Li, *Adv. Energy Mater.*, 2021, **11**, 2003002.

11  M. Ghasemi, H. Hu, Z. Peng, J. J. Rech, I. Angunawela, J. H. Carpenter, S. J. Stuard, A. Wadsworth, I. McCulloch, W. You and H. Ade, *Joule*, 2019, **3**, 1328–1348.

12   Y. Qin, N. Balar, Z. Peng, A. Gadisa, I. Angunawela, A. Bagui, S. Kashani, J. Hou and H. Ade, *Joule*, 2021, 1–19.

13   J. Luke, E. M. Speller, A. Wadsworth, M. F. Wyatt, S. Dimitrov, H. K. H. Lee, Z. Li, W. C. Tsoi, I. McCulloch, D. Bagnis, J. R. Durrant and J. Kim, *Adv. Energy Mater.*, 2019, **9**, 1803755.

14   L. Hu, Y. Liu, L. Mao, S. Xiong, L. Sun, N. Zhao, F. Qin, Y. Jiang and Y. Zhou, *J. Mater. Chem. A*, 2018, **6**, 2273–2278.

15   Y. Wang, W. Lan, N. Li, Z. Lan, Z. Li, J. Jia and F. Zhu, *Adv. Energy Mater.*, 2019, **9**, 1900157.

16   Y. Wang, J. Han, L. Cai, N. Li, Z. Li and F. Zhu, *J. Mater. Chem. A*, 2020, **8**, 21255–21264.

17   N. Grossiord, J. M. Kroon, R. Andriessen and P. W. M. Blom, *Org. Electron.*, 2012, **13**, 432–456.

18   S.-W. Liu, C.-C. Lee, W.-C. Su, C.-H. Yuan, Y.-S. Shu, W.-C. Chang, J.-Y. Guo, C.-F. Chiu, Y.-Z. Li, T.-H. Su, K.-T. Chen, P.-C. Chang, T.-H. Yeh and Y.-H. Liu, *ACS Appl. Mater. Interfaces*, 2015, **7**, 9262–9273.

19   Q. Burlingame, B. Song, L. Ciammaruchi, G. Zanotti, J. Hankett, Z. Chen, E. A. Katz and S. R. Forrest, *Adv. Energy Mater.*, 2016, **6**, 1601094.

20   M. Hermenau, S. Schubert, H. Klumbies, J. Fahlteich, L. Müller-Meskamp, K. Leo and M. Riede, *Sol. Energy Mater. Sol. Cells*, 2012, **97**, 102–108.

21   Y. Cui, Y. Xu, H. Yao, P. Bi, L. Hong, J. Zhang, Y. Zu, T. Zhang, J. Qin, J. Ren, Z. Chen, C. He, X. Hao, Z. Wei and J. Hou, *Adv. Mater.*, 2021, **33**, 2102420.

22  C. Li, J. Zhou, J. Song, J. Xu, H. Zhang, X. Zhang, J. Guo, L. Zhu, D. Wei, G. Han, J. Min, Y. Zhang, Z. Xie, Y. Yi, H. Yan, F. Gao, F. Liu and Y. Sun, *Nat. Energy*, 2021, **6**, 605–613.

23  Y. Wang, B. Wu, Z. Wu, Z. Lan, Y. Li, M. Zhang and F. Zhu, *J. Phys. Chem. Lett.*, 2017, **8**, 5264–5271.

24  J. K. J. Van Duren, X. Yang, J. Loos, C. W. T. Bulle-Lieuwma, A. B. Sieval, J. C. Hummelen and R. A. J. Janssen, *Adv. Funct. Mater.*, 2004, **14**, 425–434.

25  V. D. Mihailetchi, L. J. A. Koster, J. C. Hummelen and P. W. M. Blom, *Phys. Rev. Lett.*, 2004, **93**, 19–22.

26  T. W. Hagler, K. Pakbaz, K. F. Voss and A. J. Heeger, *Phys. Rev. B*, 1991, **44**, 8652–8666.

27  M. B. Upama, M. Wright, B. Puthen-Veettil, N. K. Elumalai, M. A. Mahmud, D. Wang, K. H. Chan, C. Xu, F. Haque and A. Uddin, *RSC Adv.*, 2016, **6**, 103899–103904.

28  T. Gotoh, S. Nonomura, S. Hirata and S. Nitta, *Appl. Surf. Sci.*, 1997, **113**–**114**, 278–281.

29  A. Weu, J. A. Kress, F. Paulus, D. Becker-Koch, V. Lami, A. A. Bakulin and Y. Vaynzof, *ACS Appl. Energy Mater.*, 2019, **2**, 1943–1950.

30  T. H. Thomas, D. J. Harkin, A. J. Gillett, V. Lemaur, M. Nikolka, A. Sadhanala, J. M. Richter, J. Armitage, H. Chen, I. McCulloch, S. M. Menke, Y. Olivier, D. Beljonne and H. Sirringhaus, *Nat. Commun.*, 2019, **10**, 2614.

31  A. Karki, J. Vollbrecht, A. J. Gillett, S. S. Xiao, Y. Yang, Z. Peng, N. Schopp, A. L. Dixon, S. Yoon, M. Schrock, H. Ade, G. N. M. Reddy, R. H. Friend and T.-Q.

Nguyen, *Energy Environ. Sci.*, 2020, **13**, 3679–3692.

32  A. Karki, J. Vollbrecht, A. J. Gillett, P. Selter, J. Lee, Z. Peng, N. Schopp, A. L. Dixon, M. Schrock, V. Nádaždy, F. Schauer, H. Ade, B. F. Chmelka, G. C. Bazan, R. H. Friend and T. Nguyen, *Adv. Energy Mater.*, 2020, **10**, 2001203.

33  S. Karuthedath, J. Gorenflot, Y. Firdaus, N. Chaturvedi, C. S. P. De Castro, G. T. Harrison, J. I. Khan, A. Markina, A. H. Balawi, T. A. Dela Peña, W. Liu, R.-Z. Liang, A. Sharma, S. H. K. Paleti, W. Zhang, Y. Lin, E. Alarousu, D. H. Anjum, P. M. Beaujuge, S. De Wolf, I. McCulloch, T. D. Anthopoulos, D. Baran, D. Andrienko and F. Laquai, *Nat. Mater.*, 2021, **20**, 378–384.

34  J. Wu, J. Lee, Y.-C. Chin, H. Yao, H. Cha, J. Luke, J. Hou, J.-S. Kim and J. R. Durrant, *Energy Environ. Sci.*, 2020, **6**, 11–13.

35  R. Wang, C. Zhang, Q. Li, Z. Zhang, X. Wang and M. Xiao, *J. Am. Chem. Soc.*, 2020, **142**, 12751–12759.

36  S. Gelinas, A. Rao, A. Kumar, S. L. Smith, A. W. Chin, J. Clark, T. S. van der Poll, G. C. Bazan and R. H. Friend, *Science (80-. ).*, 2014, **343**, 512–516.

37  A. C. Jakowetz, M. L. Böhm, A. Sadhanala, S. Huettner, A. Rao and R. H. Friend, *Nat. Mater.*, 2017, **16**, 551–557.

38  T. F. Hinrichsen, C. C. S. Chan, C. Ma, D. Paleček, A. Gillett, S. Chen, X. Zou, G. Zhang, H.-L. Yip, K. S. Wong, R. H. Friend, H. Yan, A. Rao and P. C. Y. Chow, *Nat. Commun.*, 2020, **11**, 5617.

39  A. J. Gillett, A. Privitera, R. Dilmurat, A. Karki, D. Qian, A. Pershin, G. Londi, W. K. Myers, J. Lee, J. Yuan, S.-J. Ko, M. K. Riede, F. Gao, G. C. Bazan, A. Rao, T.-Q. Nguyen, D. Beljonne and R. H. Friend, *Nature*, 2021, **597**, 666–671.

40   A. Rao, P. C. Y. Chow, S. Gélinas, C. W. Schlenker, C. Li, H. Yip, A. K.-Y. Jen, D. S. Ginger and R. H. Friend, *Nature*, 2013, **500**, 435–439.

41   I. A. Howard, R. Mauer, M. Meister and F. Laquai, *J. Am. Chem. Soc.*, 2010, **132**, 14866–14876.

42   F. Deschler, A. De Sio, E. von Hauff, P. Kutka, T. Sauermann, H.-J. Egelhaaf, J. Hauch and E. Da Como, *Adv. Funct. Mater.*, 2012, **22**, 1461–1469.

43   I. Ramirez, A. Privitera, S. Karuthedath, A. Jungbluth, J. Benduhn, A. Sperlich, D. Spoltore, K. Vandewal, F. Laquai and M. Riede, *Nat. Commun.*, 2021, **12**, 471.

44   K. Wang, H. Chen, S. Li, J. Zhang, Y. Zou and Y. Yang, *J. Phys. Chem. B*, 2021, **125**, 7470–7476.

45   T. H. Thomas, J. P. H. Rivett, Q. Gu, D. J. Harkin, J. M. Richter, A. Sadhanala, C. K. Yong, S. Schott, K. Broch, J. Armitage, A. J. Gillett, S. M. Menke, A. Rao, D. Credgington and H. Sirringhaus, *ACS Nano*, 2019, **13**, 13716–13727.

46   M. Yan, L. J. Rothberg, F. Papadimitrakopoulos, M. E. Galvin and T. M. Miller, *Phys. Rev. Lett.*, 1994, **72**, 1104–1107.

47   E. L. Frankevich, A. A. Lymarev, I. Sokolik, F. E. Karasz, S. Blumstengel, R. H. Baughman and H. H. Hörhold, *Phys. Rev. B*, 1992, **46**, 9320–9324.

48   S. M. Menke, A. Cheminal, P. Conaghan, N. A. Ran, N. C. Greehnam, G. C. Bazan, T.-Q. Nguyen, A. Rao and R. H. Friend, *Nat. Commun.*, 2018, **9**, 277.

49   J. Niklas and O. G. Poluektov, *Adv. Energy Mater.*, 2017, **7**, 1602226.

50   J. Niklas, S. Beaupré, M. Leclerc, T. Xu, L. Yu, A. Sperlich, V. Dyakonov and O. G. Poluektov, *J. Phys. Chem. B*, 2015, **119**, 7407–7416.

51  M. Righetto, A. Privitera, F. Carraro, L. Bolzonello, C. Ferrante, L. Franco and R. Bozio, *Nanoscale*, 2018, **10**, 11913–11922.

52  H. Matsui, T. Hasegawa, Y. Tokura, M. Hiraoka and T. Yamada, *Phys. Rev. Lett.*, 2008, **100**, 126601.

53  C. Carati, L. Bonoldi and R. Po, *Phys. Rev. B*, 2011, **84**, 245205.

54  N. Camaioni, F. Tinti, L. Franco, M. Fabris, A. Toffoletti, M. Ruzzi, L. Montanari, L. Bonoldi, A. Pellegrino, A. Calabrese and R. Po, *Org. Electron.*, 2012, **13**, 550–559.

55  K. Marumoto, Y. Muramatsu and S. Kuroda, *Appl. Phys. Lett.*, 2004, **84**, 1317–1319.

56  E. A. Lukina, M. N. Uvarov and L. V. Kulik, *J. Phys. Chem. C*, 2014, **118**, 18307–18314.

57  W. Li, D. Liu and T. Wang, *Adv. Funct. Mater.*, 2021, **31**, 2104552.

58  Y. Wu, Y. Zheng, H. Yang, C. Sun, Y. Dong, C. Cui, H. Yan and Y. Li, *Sci. China Chem.*, 2020, **63**, 265–271.

59  J. C. de Mello, H. F. Wittmann and R. H. Friend, *Adv. Mater.*, 1997, **9**, 230–232.

60  S. Stoll and A. Schweiger, *J. Magn. Reson.*, 2006, **178**, 42–55.

61  M. J. . T. Frisch G. W.; Schlegel, H. B.; Scuseria, G. E.; Robb, M. A.; Cheeseman, J. R.; Scalmani, G.; Barone, V.; Petersson, G. A.; Nakatsuji, H.; Li, X.; Caricato, M.; Marenich, A. V.; Bloino, J.; Janesko, B. G.; Gomperts, R.; Mennucci, B.; Hratch, D. J., *Gaussian, Inc., Wallingford, CT*.

62  V. A. Tran and F. Neese, *J. Chem. Phys.*, 2020, **153**, 054105.

63  F. Neese, F. Wennmohs, U. Becker and C. Riplinger, *J. Chem. Phys.*, 2020, **152**, 224108.

# Supplementary Information for

# The critical role of the donor polymer in the stability of high-performance non-fullerene acceptor organic solar cells

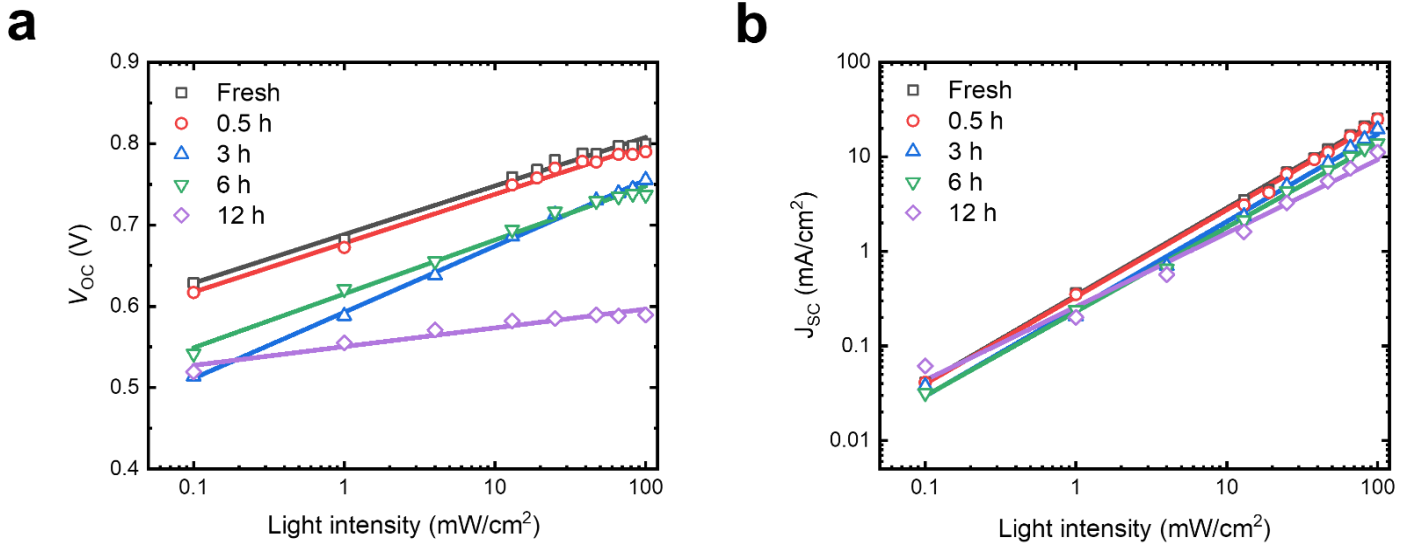

**Figure S1:** (a) $V_{OC}$–ln $I$ plots of the fresh PM6:Y6 devices and samples light soaked for up to 12 hours in ambient air. (b) $J_{SC}$–ln $I$ plots of the fresh PM6:Y6 devices and samples light soaked for up to 12 hours in ambient air.

| Light soaking time | $V_{OC}$–ln $I$ slope ($\frac{kT}{q}$) | $J_{SC}$–ln $I$ |
|---|---|---|
| Fresh | 1.00 | 0.924 |
| 0.5 hours | 1.00 | 0.919 |
| 3 hours | 1.37 | 0.924 |
| 6 hours | 1.11 | 0.891 |
| 12 hours | 0.39 | 0.777 |

**Table S1:** A summary of the gradients extracted from the light intensity dependent device studies.

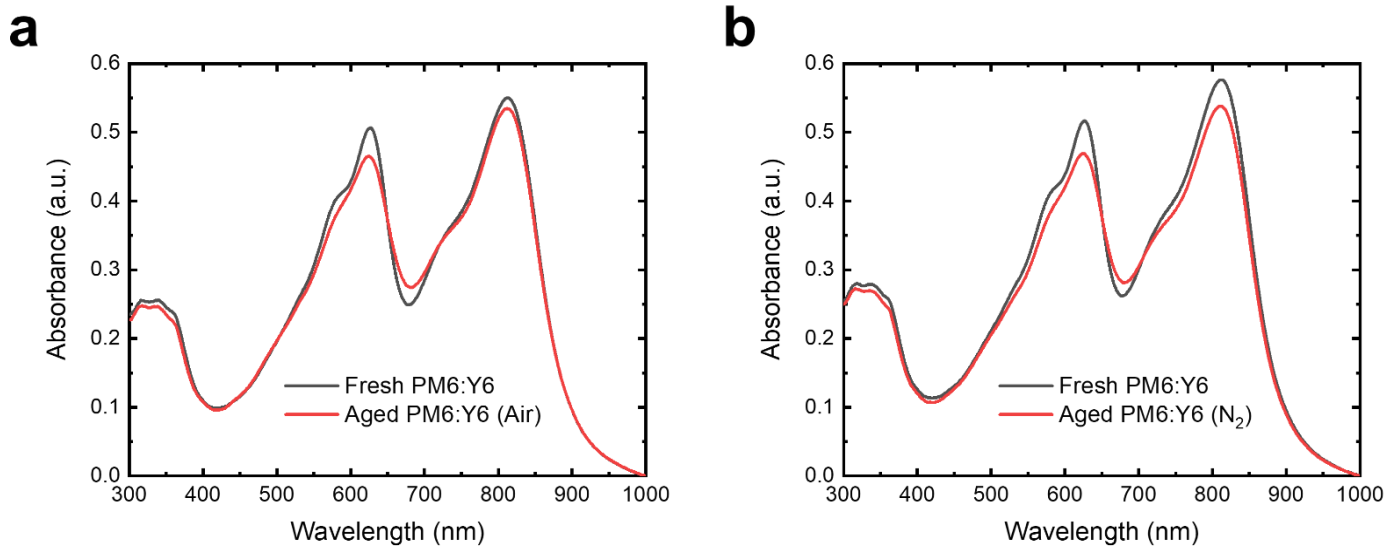

**Figure S2:** **(a)** The UV-vis absorption spectrum of a fresh PM6:Y6 film and a sample light-soaked for 12 hours in air. **(b)** The UV-vis absorption spectrum of a fresh PM6:Y6 film and a sample light-soaked for 12 hours in $N_2$.

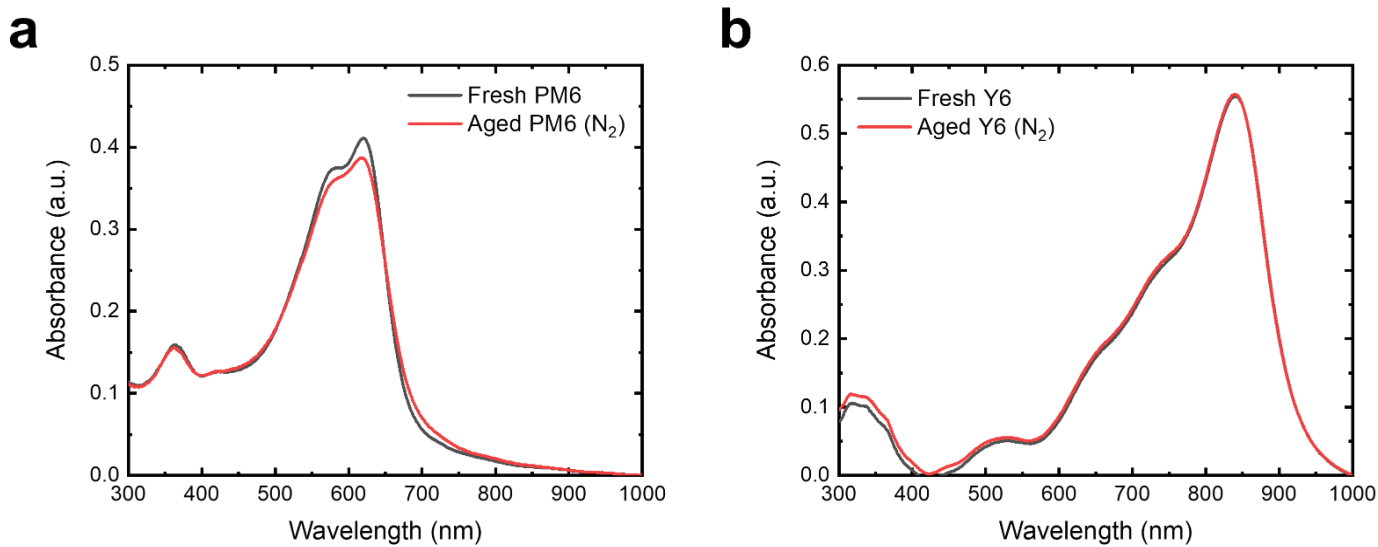

**Figure S3:** **(a)** The UV-vis absorption spectrum of a fresh PM6 film and a sample light-soaked for 12 hours in $N_2$. **(b)** The UV-vis absorption spectrum of a fresh Y6 film and a sample light-soaked for 12 hours in $N_2$.

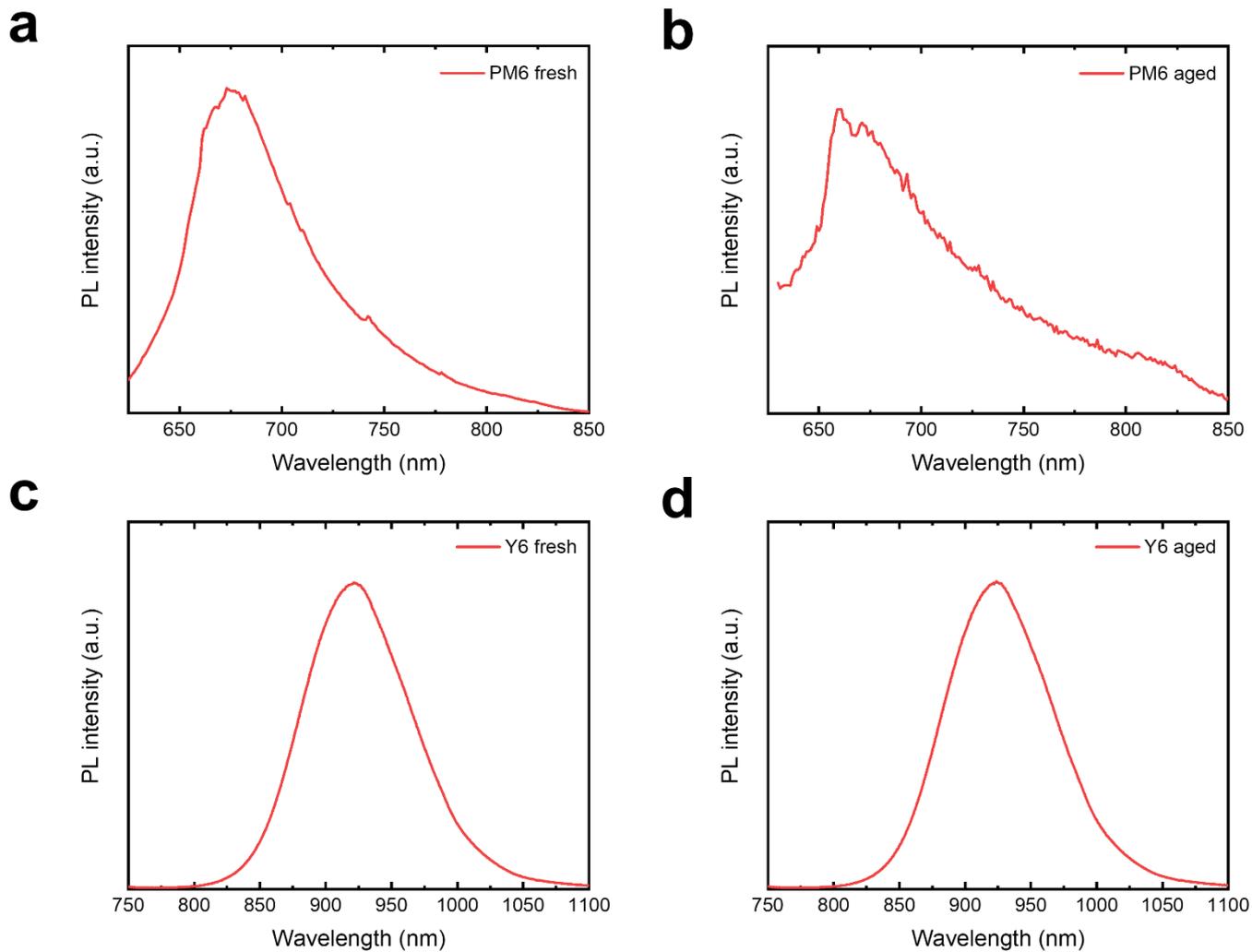

**Figure S4:** **(a)** The PL spectrum of a fresh PM6 film. **(b)** The PL spectrum of a PM6 film light-soaked for 12 hours in air. **(c)** The PL spectrum of a fresh Y6 film. **(d)** The PL spectrum of a Y6 film light-soaked for 12 hours in air.

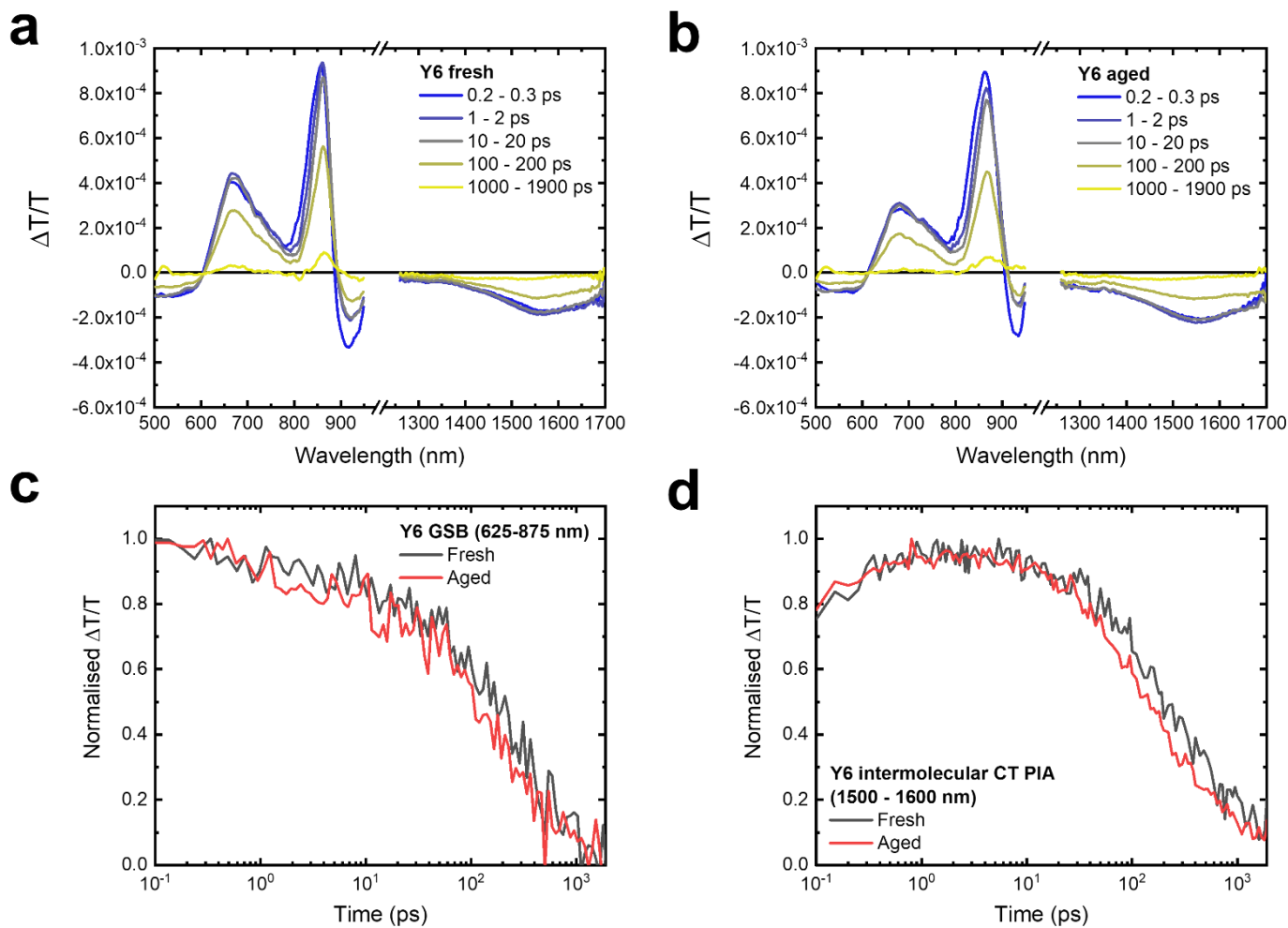

**Figure S5:** **(a)** The TA spectra of the fresh neat Y6 film, excited at 800 nm with a fluence of 0.88 μJ/cm$^2$. **(b)** The TA spectra of the neat Y6 film, light-soaked overnight for 12 hours in air, excited at 800 nm with a fluence of 0.88 μJ/cm$^2$. **(c)** The normalized kinetics of the Y6 GSB region (625-875 nm), taken from the fresh and aged neat Y6 films. **(d)** The normalized kinetics of the Y6 intermolecular CT state PIA (1500-1600 nm), taken from the fresh and aged neat Y6 films.

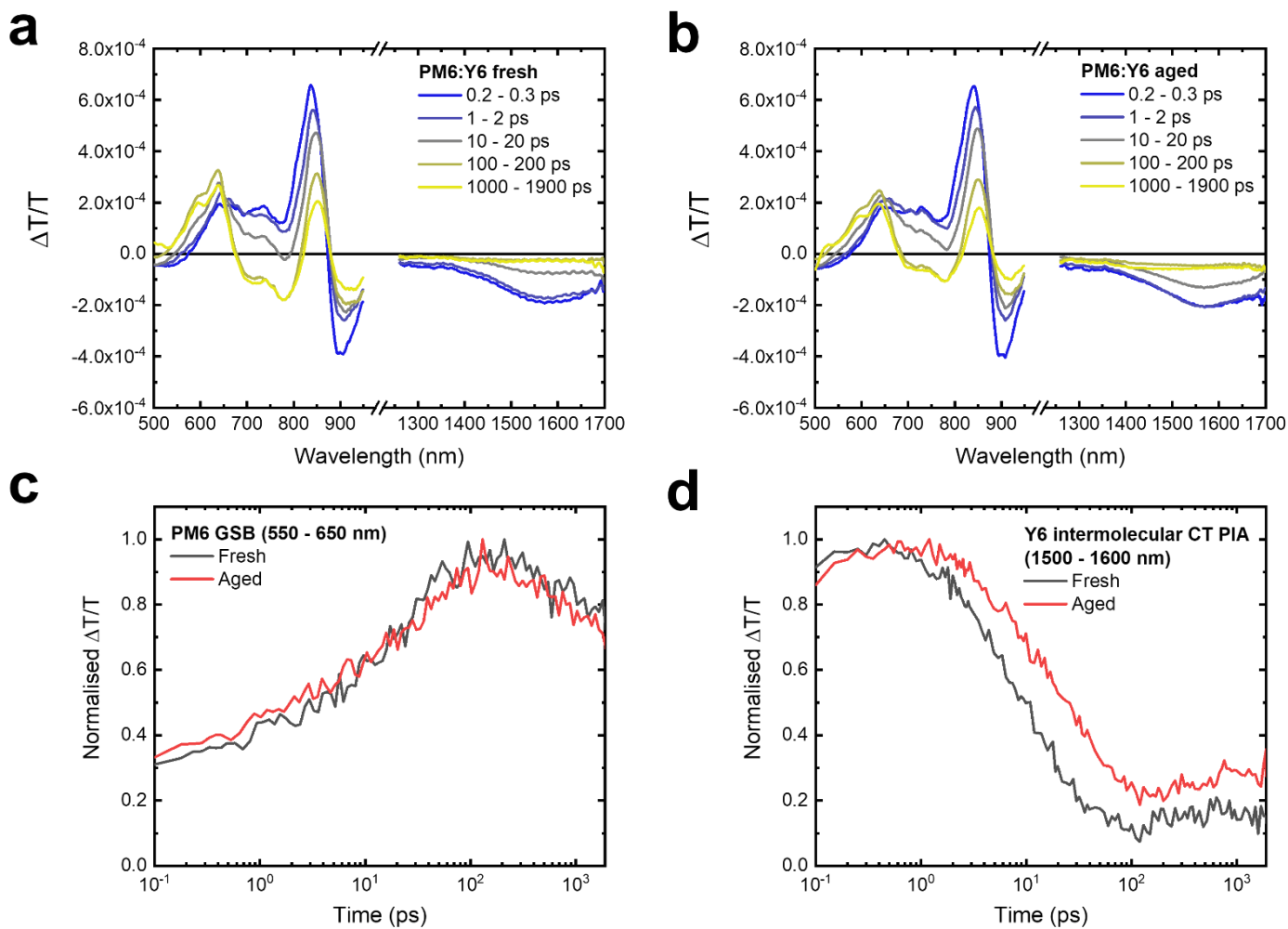

**Figure S6:** (**a**) The TA spectra of the fresh PM6:Y6 blend film, excited at 800 nm with a fluence of 0.55 μJ/cm$^2$. (**b**) The TA spectra of the PM6:Y6 blend film, light-soaked for 12 hours in air, excited at 800 nm with a fluence of 0.55 μJ/cm$^2$. The intensity of the PM6 GSB and hole polaron PIA, after charge transfer is completed by 1000-1900 ps, are slightly lower in the aged film compared to the fresh sample. This indicates a minor reduction in the charge generation yield in the aged PM6:Y6 films from excitons generated on Y6. (**c**) The normalized kinetics of the PM6 GSB region (550-650 nm), taken from the fresh and aged PM6:Y6 blend films. The near identical kinetics show that the interfacial hole transfer rate is the same in the fresh and aged PM6:Y6 films. (**d**) The raw kinetics of the Y6 intermolecular CT state PIA (1500-1600 nm), taken from the fresh and aged PM6:Y6 blend films. The intermolecular CT state that mediates hole transfer from Y6 decays more slowly in the aged blend, attributed to the increased domain sizes in PM6 and Y6 resulting from phase separation upon aging.

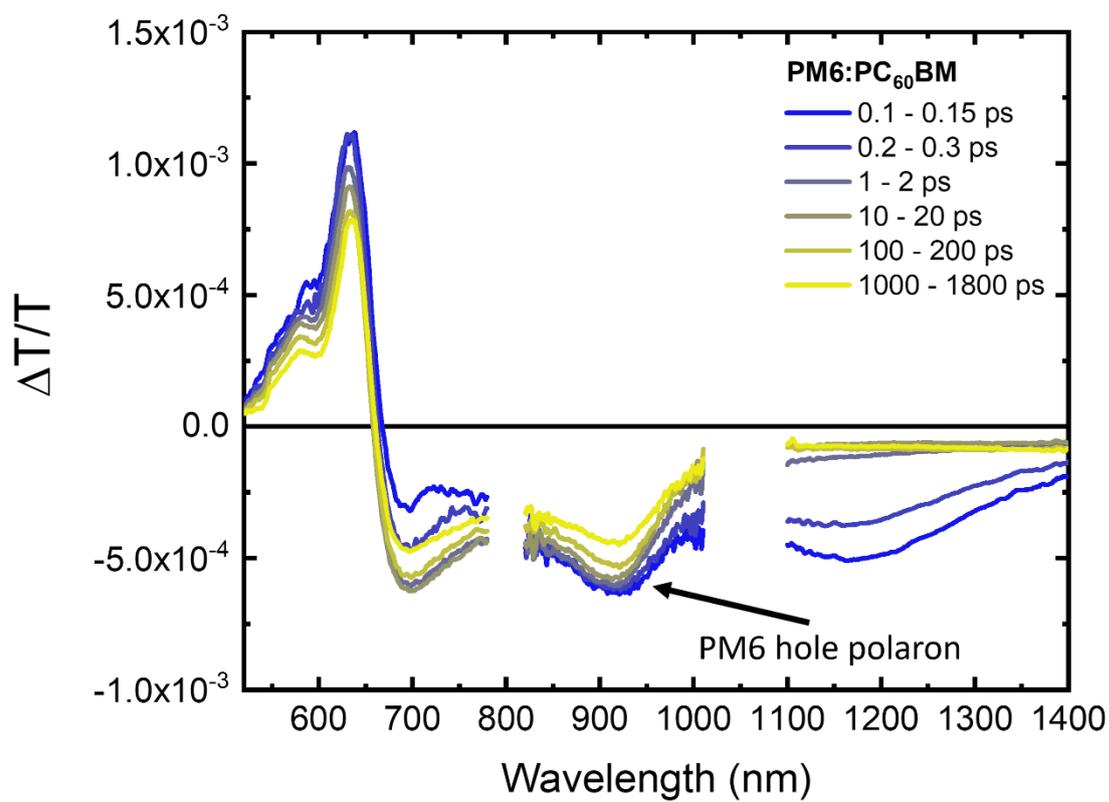

**Figure S7:** The TA spectra of a PM6:PC$_{60}$BM blend film, excited at 580 nm with a fluence of 0.62 μJ/cm$^2$. The PM6 hole polaron PIA is at 920 nm.

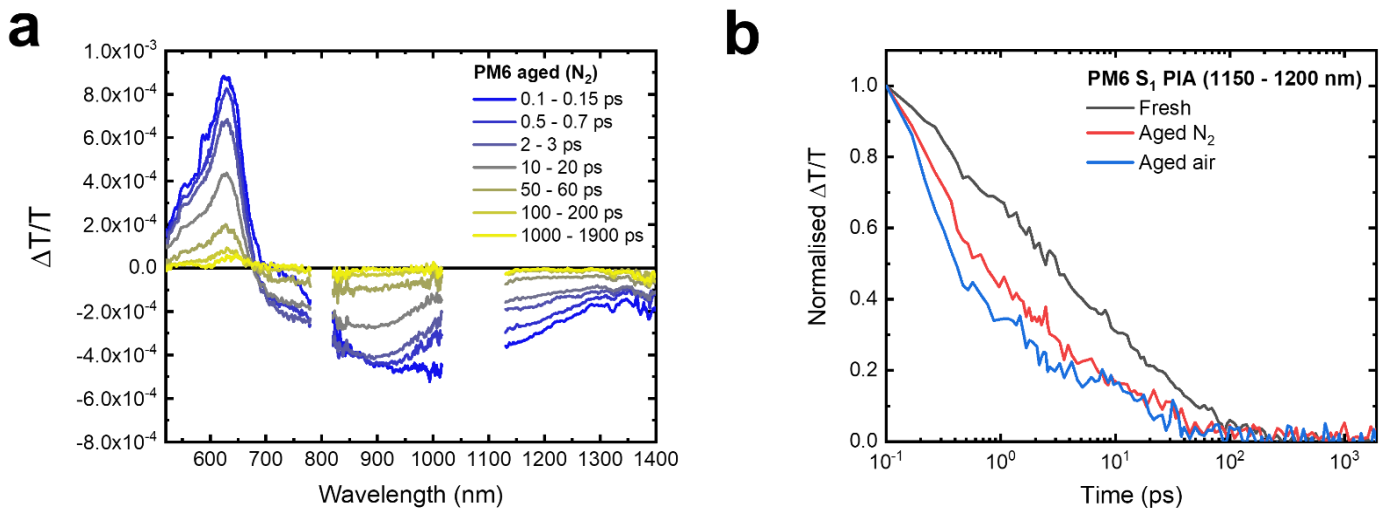

**Figure S8: (a)** The TA spectra of the neat PM6 film, light-soaked for 12 hours in an $N_2$ environment, excited at 580 nm with a fluence of 0.64 μJ/cm². **(b)** The normalized kinetics of the PM6 $S_1$ PIA (1150-1200 nm), taken from the fresh PM6 film and the samples light-soaked for 12 hours in $N_2$ and ambient air. The increased PM6 $S_1$ PIA quenching is very similar in the $N_2$ and air aged films, indicating that this process is primarily driven by light soaking.

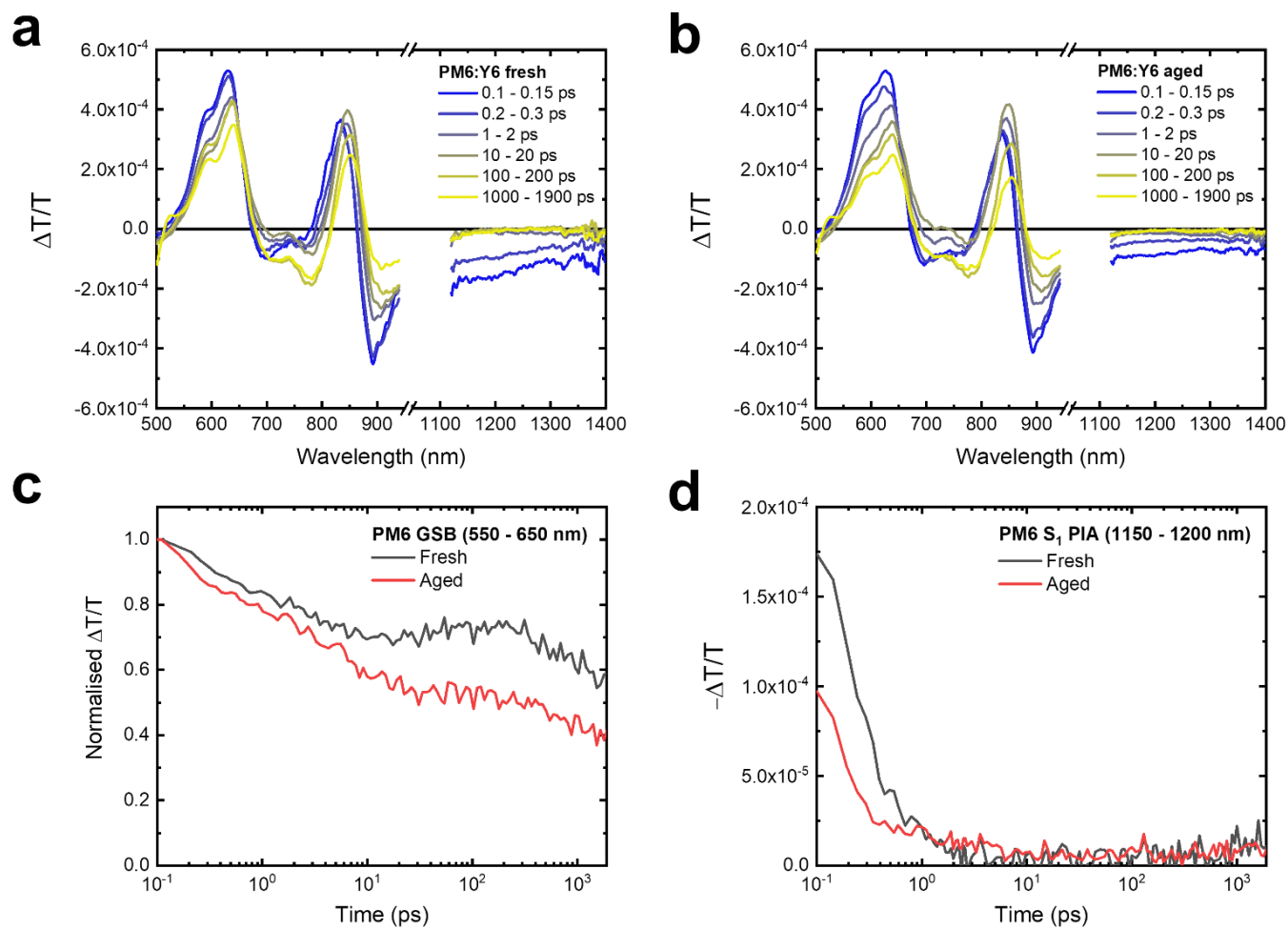

**Figure S9: The TA spectra and kinetics of fresh and aged PM6:Y6 blend films.** (**a**) The TA spectra of the fresh PM6:Y6 blend film, excited at 580 nm with a fluence of 0.64 μJ/cm$^2$. (**b**) The TA spectra of the PM6:Y6 blend film, light soaked for 12 hours in ambient air, excited at 580 nm with a fluence of 0.64 μJ/cm$^2$. (**c**) The normalized kinetics of the PM6 GSB region (550-650 nm), taken from the fresh and aged PM6:Y6 blend films. (**d**) The raw kinetics of the PM6 S$_1$ PIA (1150-1200 nm), taken from the fresh and aged PM6:Y6 blend films.

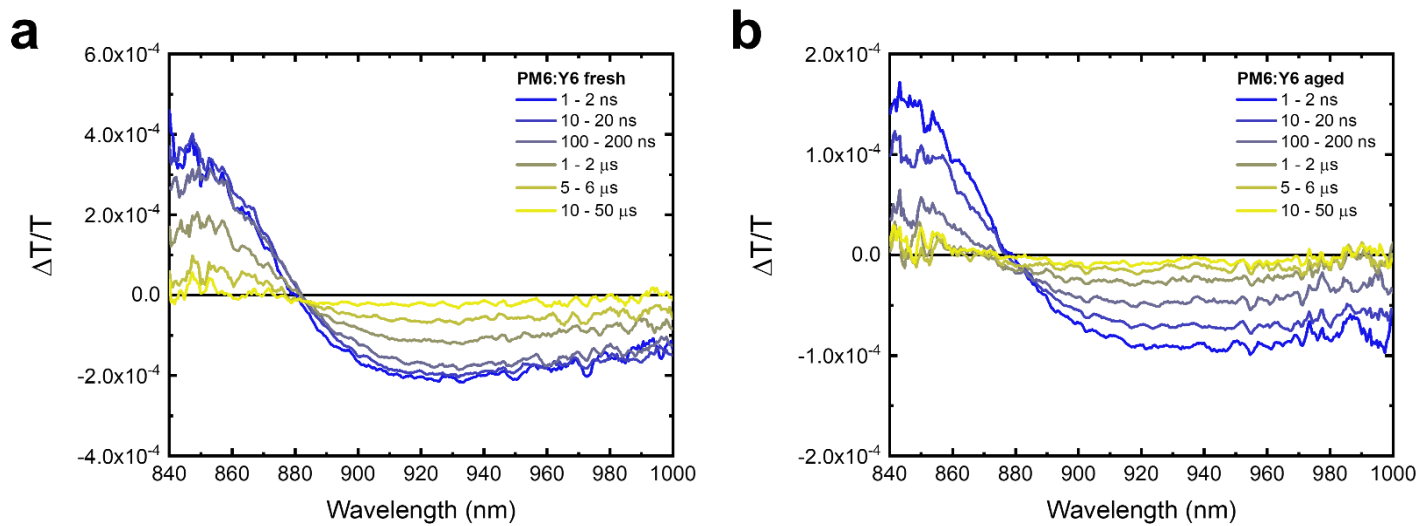

**Figure S10:** (a) The nanosecond TA spectra of the fresh PM6:Y6 film, excited at 532 nm with a fluence of 0.20 μJ/cm$^2$. (b) The nanosecond TA spectra of the PM6:Y6 film, light-soaked in air for 12 hours, excited at 532 nm with a fluence of 0.20 μJ/cm$^{-2}$.

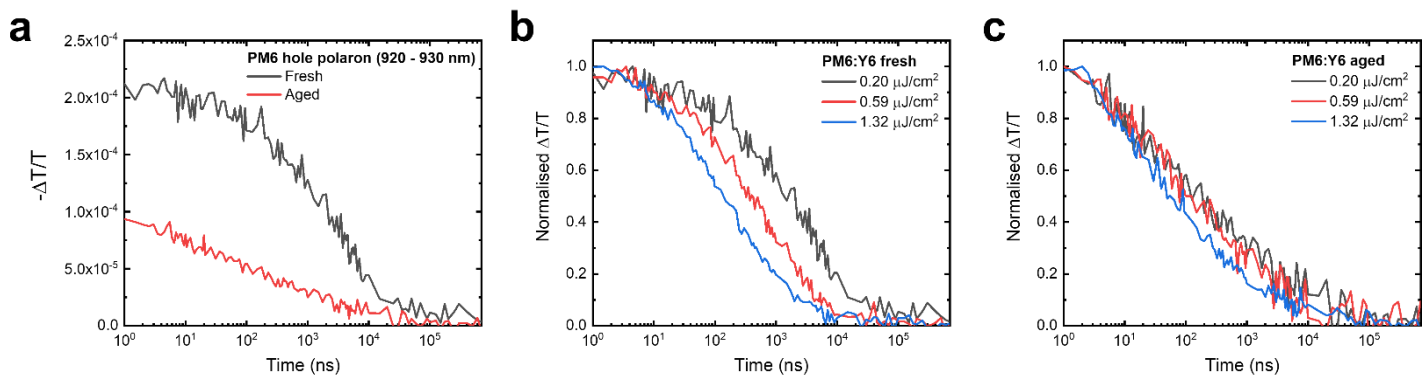

**Figure S11: The nanosecond TA kinetics of fresh and aged PM6:Y6 blend films.** (a) The raw kinetics of the fresh and aged PM6:Y6 blend films, taken from the PM6 hole polaron PIA between 920-930 nm. The films were excited at 532 nm with a fluence of 0.2 μJ/cm$^2$. (b) The normalized fluence series kinetics of the fresh PM6:Y6 blend film after 532 nm excitation, taken from the PM6 hole polaron PIA between 920-930 nm. (c) The normalized fluence series kinetics of the PM6:Y6 blend film, light soaked for 12 hours in ambient air, after 532 nm excitation. Kinetics taken from the PM6 hole polaron PIA between 920-930 nm.

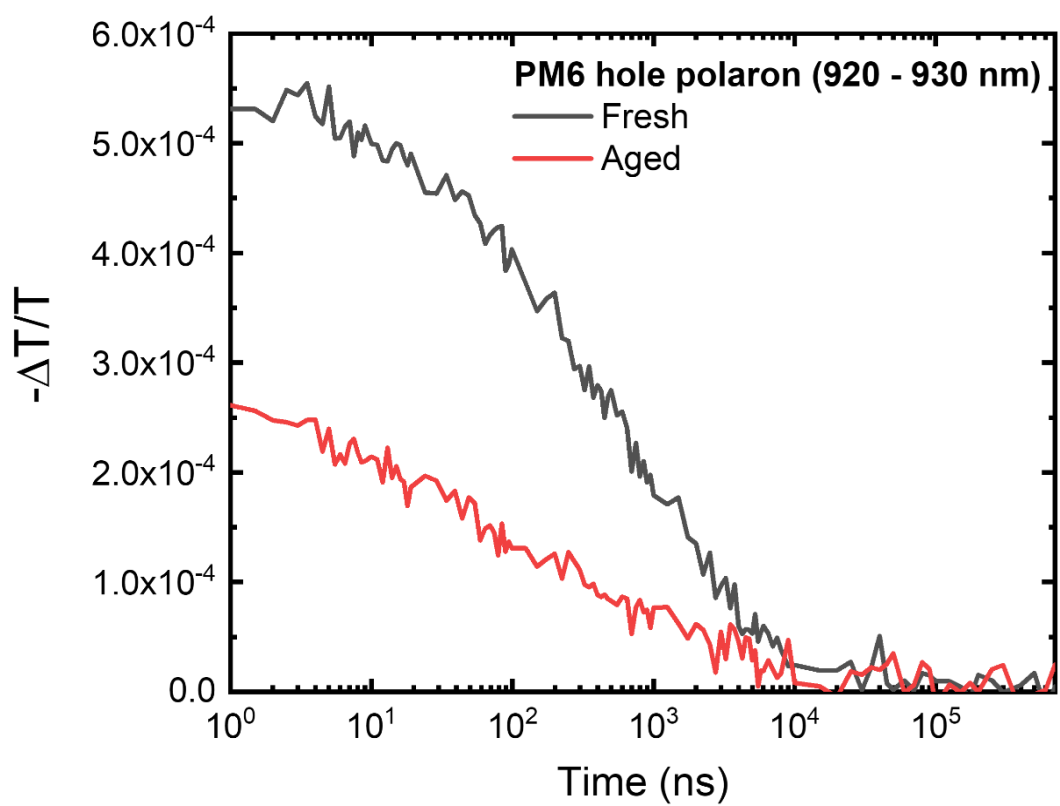

**Figure S12:** The raw kinetics of the fresh PM6:Y6 blend film and the sample light-soaked in air for 12 hours, taken from the PM6 hole polaron PIA between 920-930 nm. The films were excited at 532 nm with a fluence of 0.59 μJ/cm$^2$. Consistent with the lower fluence (0.2 μJ/cm$^2$) measurement, there is roughly a factor of two difference in the intensity of the hole polaron PIA between the fresh and aged films.

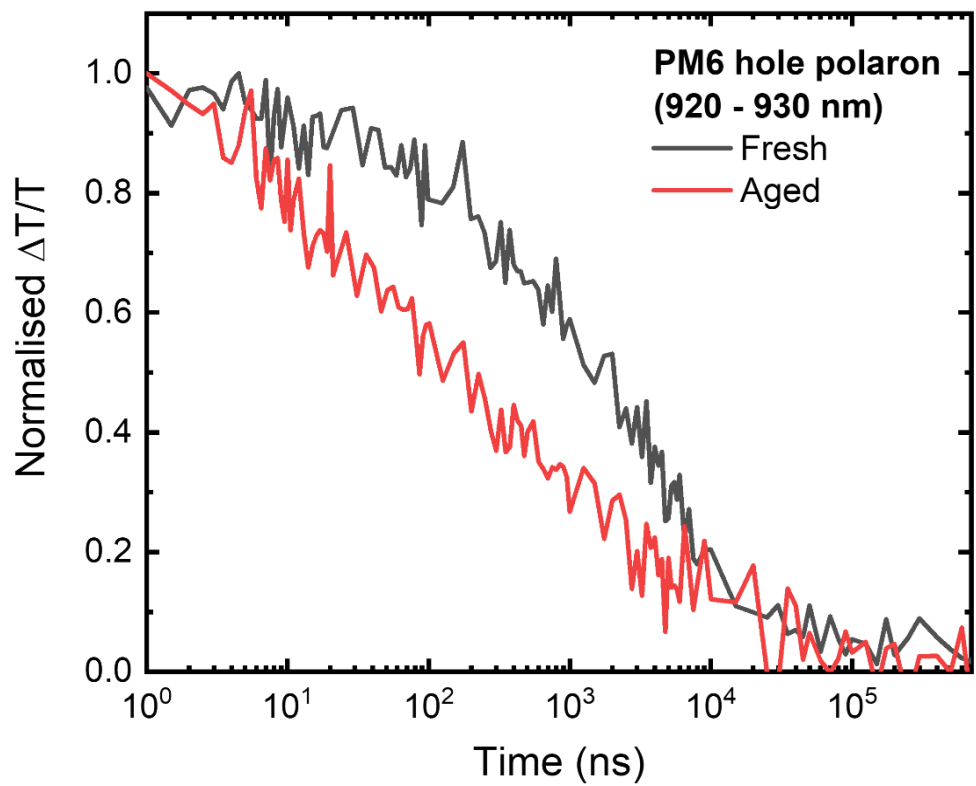

**Figure S13:** The normalized kinetics of the fresh PM6:Y6 blend film and the sample light-soaked in air for 12 hours, taken from the PM6 hole polaron PIA between 920-930 nm. The films were excited at 532 nm with a fluence of 0.20 µJ/cm$^2$. Despite the hole population being roughly twice as large in the fresh film at 1 ns, the recombination is significantly faster in the aged blend.

**Discussion of PM6:Y6 blend TA**

We now explore the impact of the ultrafast interchain PP formation in aged PM6 on the blend photophysics. To preferentially create excitons on PM6, we have excited the PM6:Y6 blend at 580 nm where there is significant PM6 absorption and weaker absorption by Y6. At 0.1-0.15 ps after excitation in the TA of the fresh blend (Figure S9a), we observe the PM6 ground GSB between 550-650 nm and the PM6 $S_1$ PIA at 1150 nm. The Y6 GSB, peaked at 840 nm, is also present at this early time, largely formed *via* ultrafast electron transfer from PM6 to Y6, though the Y6 $S_1$ PIA at 900 nm also confirms some direct photoexcitation of Y6. Within 1 ps, the PM6 $S_1$ PIA is fully quenched by electron transfer to Y6 (Figure S9d). Following this, we observe a decrease in the PM6 GSB intensity up to 10 ps, accompanied by a small red shift of this feature, which we tentatively attribute to relaxation of the hole polarons created on the polymer chain after electron transfer[1]. After 10 ps, the PM6 GSB rises again, consistent with the timescales of hole transfer from Y6 to PM6; this is caused by the smaller fraction of excitons directly generated on Y6. Finally, we note the presence of the long lived PM6 hole polaron PIA at 920 nm.

In contrast, the PM6 GSB kinetic in the aged blend begins to deviate strongly from the fresh blend at around 10 ps (Figure S9c), suggesting an additional process is quenching the electronic excitations present on PM6. In addition, despite exciting both blends with the same fluence, the intensity of the PM6 $S_1$ PIA at the earliest time resolvable (0.1-0.15 ps) in the aged blend is significantly weaker (Figures S9b, S9d). As the rate of the interfacial hole transfer process from Y6 to PM6 does not change in the aged film (Figure S6c), which implies that the D:A interfacial morphology is not significantly affected upon aging, we would also anticipate that the interfacial electron transfer rate is largely unchanged. Indeed, the larger domains formed from the D:A de-mixing should, if anything, result in slower PM6 $S_1$ PIA quenching due to the increased exciton diffusion distance to the D:A interface[2]. Thus, we conclude that the ultrafast PP formation in aged PM6 is fast enough to out-compete electron transfer to Y6, accounting for the reduced PM6 $S_1$ PIA intensity at 0.1-0.15 ps. It has also been suggested that since the interchain PPs in PM6 have a much shorter lifetime and slower transport though the PM6 domains than $S_1$ states, they are susceptible to recombining before reaching the D:A interface for electron transfer[3]. Therefore, as the PM6 $S_1$ states are more rapidly funnelled through this channel in the aged samples, this results in increased losses though interchain PP recombination. This is supported by TA of the fresh and aged PM6:Y6 blends with 580 nm excitation, where the divergence in the PM6 GSB kinetics takes place on the same tens of picoseconds timescales that the interchain PP states recombine over (Figures 3c, S9c).

Having now identified a key loss channel intrinsic to PM6, we next evaluate its effect on the charge generation efficiency of the blend. To achieve this, we have performed nanosecond TA studies on the fresh and aged blends, excited at 532 nm for preferential PM6 excitation (Figure S10). In Figure S11a, we present the raw recombination kinetics of the PM6 hole polaron in the blends, taken between 920-930 nm (corresponding TA spectra in Figure S8). For these measurements, an extremely low excitation fluence of 0.20 µJ/cm$^2$ was used for both samples to ensure no excess bimolecular recombination occurred on sub-nanosecond timescales. We observe that the intensity of the PM6 hole polaron signal at 1 ns in the aged blend is about half that of the fresh blend, confirming that the charge yield following the electron transfer process is severely limited by the PM6 interchain PP loss pathway. The difference in the hole polaron yields also agrees well with the roughly factor of two lower intensity of the PM6 $S_1$ PIA at 0.1-0.15 ps (Figures S9a, S9b), indicating that ultrafast interchain PP formation is the primary photocurrent loss channel in the aged blend.

Measurements at a higher fluence of 0.59 µJ/cm$^2$ corroborate this factor of two change in the hole polaron yield (Figure S12), reinforcing that this difference does not result from sub-nanosecond bimolecular recombination events. Furthermore, the normalised hole polaron kinetics show that the charge recombination is significantly faster in the aged blend (Figure S13), despite the factor of two lower charge density at 1 ns. A fluence series reveals a strong fluence dependence in the hole polaron decay in the fresh blend (Figure S9b), indicative of dominant bimolecular recombination processes[1,4]. In contrast, the aged blend shows little fluence dependence to the hole polaron decay (Figure S9c), consistent with the observation of increased trap-mediated recombination processes in the aged PM6:Y6 devices (Figure S1a).

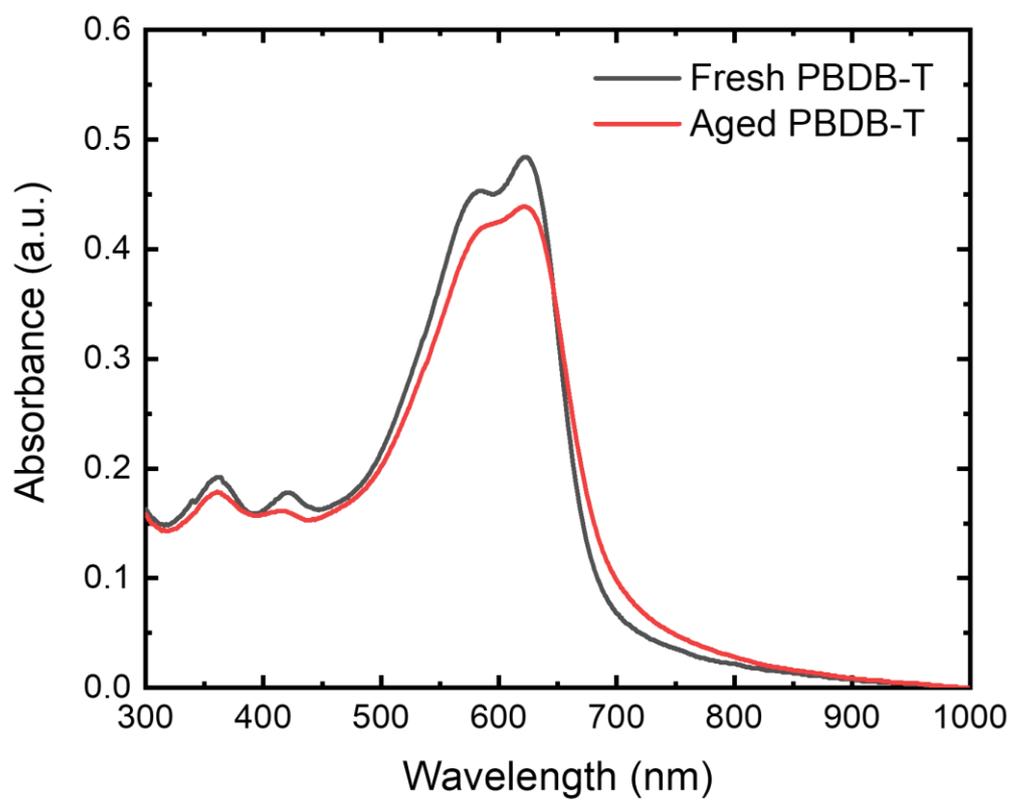

**Figure S14:** The UV-vis absorption spectrum of a fresh PBDB-T film and a sample light-soaked for 12 hours in air. Like PM6, there is a loss of a vibronic structure of the absorption band and an increase in the sub gap absorption tail upon aging.

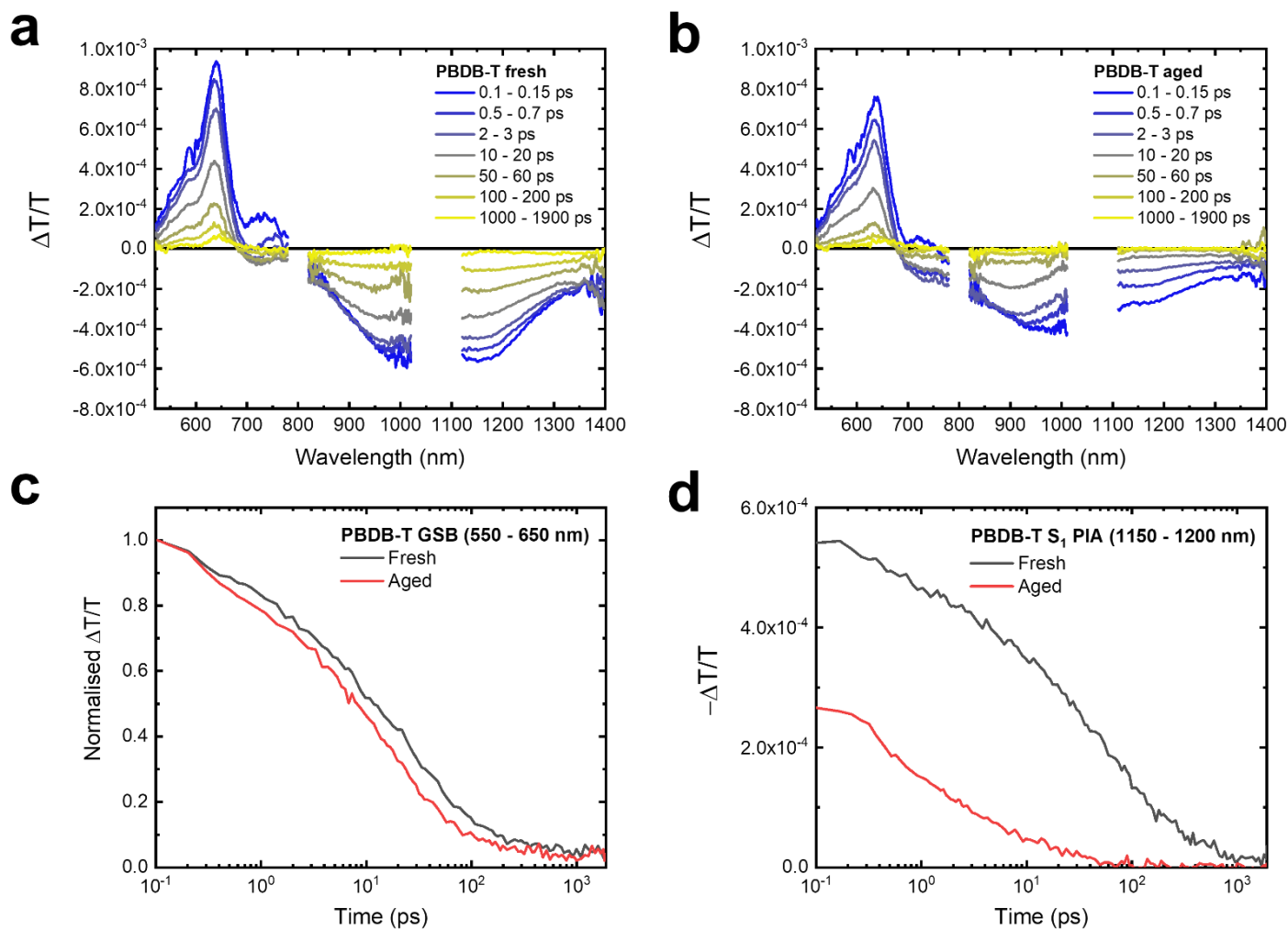

**Figure S15:** **(a)** The TA spectra of the fresh neat PBDB-T film, excited at 600 nm with a fluence of 0.64 μJ/cm$^2$. **(b)** The TA spectra of the neat PBDB-T film, light soaked for 12 hours in ambient air, excited at 600 nm with a fluence of 0.64 μJ/cm$^2$. **(c)** The normalized kinetics of the PBDB-T GSB region (550-650 nm), taken from the fresh and aged neat PBDB-T films. **(d)** The raw kinetics of the PBDB-T S$_1$ PIA (1150-1200 nm), taken from the fresh and aged PBDB-T films.

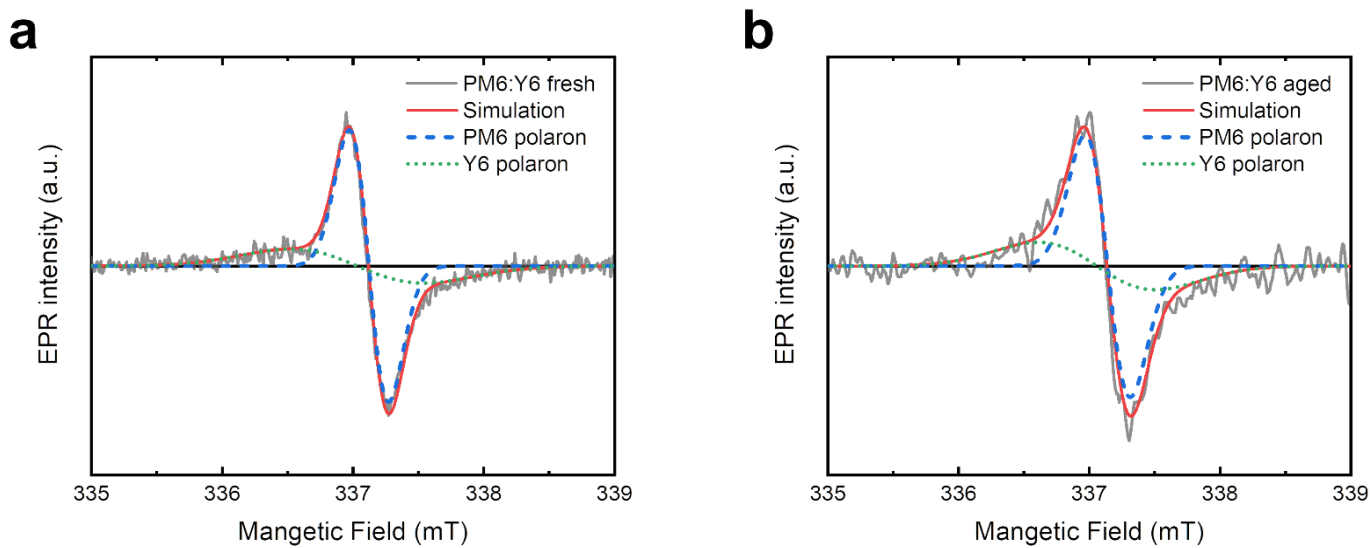

**Figure S16:** (**a**) Light-induced EPR spectra and best-fit simulation of the fresh PM6:Y6 blend at 5 K, excited at 534 nm. (**b**) Light-induced EPR spectra and best-fit simulation of the aged PM6:Y6 blend at 10 K, excited at 534 nm.

The experimental spectra (black line) shown are the difference between spectra recorded under illumination of the sample at 534 nm (light on) and those recorded before sample illumination (dark). The best-fit simulations (red lines) are the sum of two species (blue dashed and green dotted lines) characterized by two different g-values. Simulation parameters are given in Table S2. Quantum chemical calculations of the PM6 and Y6 polaron g-values enable us to attribute the feature at g~2.0028 to the positive polaron on PM6 and the feature at g~2.0033 to the negative polaron on Y6.

| Sample | Polaron 1 g-value | Polaron 1 linewidth (mT) | Polaron 1 spectral weighting | Polaron 2 g-value | Polaron 2 linewidth (mT) | Polaron 2 spectral weighting |
|---|---|---|---|---|---|---|
| **PM6:Y6 fresh** | 2.0028 (PM6) | 0.29 | 0.41 | 2.0033 (Y6) | 1.0 | 0.59 |
| **PM6:Y6 aged** | 2.0027 (PM6) | 0.35 | 0.45 | 2.0031 (Y6) | 0.9 | 0.55 |

**Table S2:** Simulation parameters of the best-fit L-EPR simulations of the fresh and aged PM6:Y6 blends at 10 K.

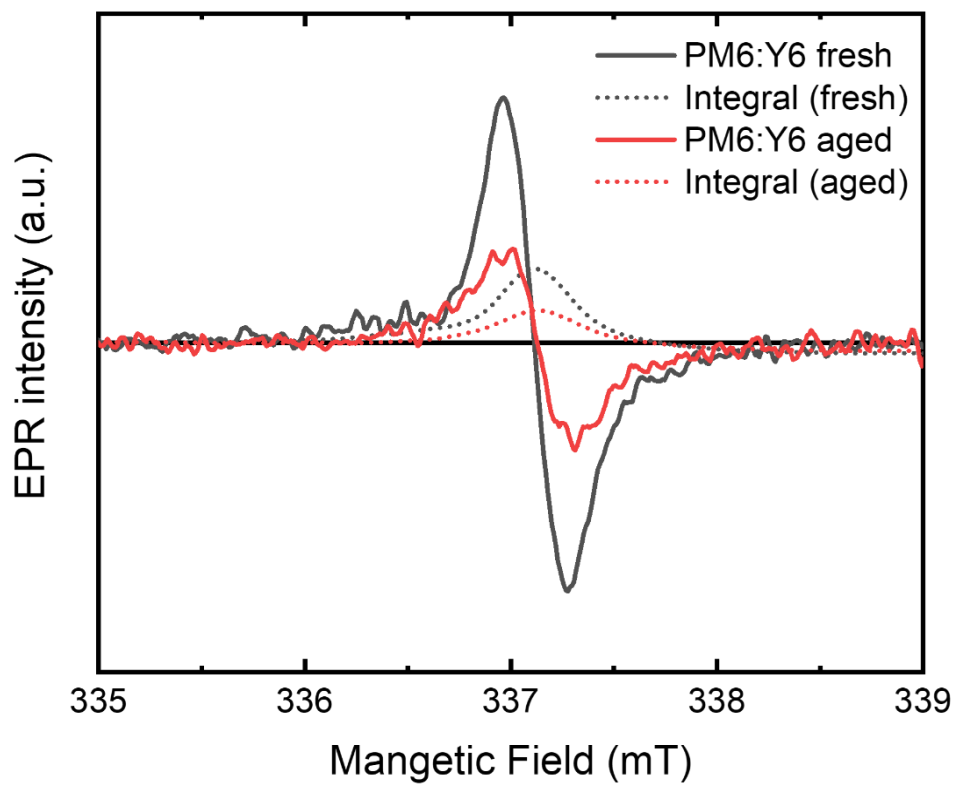

**Figure S17:** Integrated L-EPR spectra of fresh and aged PM6:Y6 blends at 5 K. The spectra shown are the difference between spectra recorded under illumination of the sample at 534 nm (light on) and those recorded before sample illumination (dark). The studied samples are the fresh (red line) and aged (black line) PM6:Y6 blend. The integrated spectra are reported with dotted lines. From the double integration, we obtained an area of 603 and 313 respectively (a.u.) for the fresh and the aged blends, which indicates that the L-EPR signal of the fresh blend is twice as large as the aged sample.

## Supplementary References

1   F. Etzold, I. A. Howard, R. Mauer, M. Meister, T.-D. Kim, K.-S. Lee, N. S. Baek and F. Laquai, *J. Am. Chem. Soc.*, 2011, **133**, 9469–9479.

2   Y. Qin, N. Balar, Z. Peng, A. Gadisa, I. Angunawela, A. Bagui, S. Kashani, J. Hou and H. Ade, *Joule*, 2021, 1–19.

3   K. Wang, H. Chen, S. Li, J. Zhang, Y. Zou and Y. Yang, *J. Phys. Chem. B*, 2021, **125**, 7470–7476.

4   I. A. Howard, R. Mauer, M. Meister and F. Laquai, *J. Am. Chem. Soc.*, 2010, **132**, 14866–14876.